\documentclass{ws-rv975x65}
\usepackage{ws-rv-van}     
\newcommand{\Tr}{\textrm{Tr}}

\begin{document}
\vspace*{-0.3cm}
\begin{center}
{\Large \bf The Skyrme model in the BPS limit\footnote{To appear in: M. Rho, I. Zahed, Eds., {\em The Multifaceted Skyrmion}, 2nd Edition, World Scientific, Singapore}}
\end{center}
\vspace*{0.8cm}
\author[C. Adam]{C. Adam, C. Naya, J. S\'{a}nchez-Guill\'{e}n, R.  Vazquez}

\address{Departamento de F\'isica de Part\'iculas, Universidad de Santiago de Compostela \\
and 
\\
Instituto Galego de F\'isica de Altas Enerxias (IGFAE) \\
E-15782 Santiago de Compostela, Spain}

\author[]{A. Wereszczy\'{n}ski}
\address{Institute of Physics,  Jagiellonian University,
Lojasiewicza 11, Krak\'{o}w, Poland}

\vspace*{0.5cm}

\begin{abstract}
{
In this review, we summarise the main features of the BPS Skyrme model
which provides a
physically well-motivated idealisation of atomic nuclei and
nuclear matter: 
1) it leads to zero binding energies for classical solitons (while
realistic binding energies 
emerge owing to the semiclassical corrections, the Coulomb interaction and
isospin breaking); 
2) it describes a perfect non-barotropic fluid
already at the microscopic (field theoretical)
level which allows to study thermodynamics beyond the mean-field
limit.
These properties allow for an approximate but analytical calculation of binding
energies of the most abundant nuclei, for a determination of the equation of state of 
skyrmionic matter (both in the full field theory and in a mean-field approximation) as well as
the description of neutron stars as Skyrme solitons with a very good agreement with
available observational data.

All these results suggest that the proper low energy effective model of QCD
should be close to the BPS Skyrme model in a certain sense 
(a "near-BPS Skyrme model"),  with a prominent role played by the BPS part.
}
\end{abstract}
\body
{\LARGE
\tableofcontents
}
\vspace*{0.8cm}
\section{Motivation}\label{mot}
The precise derivation of low energy properties of strongly interacting matter directly from Quantum Chromodynamics (QCD) is one of the most  important unsolved problems of contemporary theoretical particle physics, related to the non-perturbative character of the  low energy sector. Still, significant insight into this regime may be gained with the help of low energy effective field theories (EFTs). Although the precise derivation of these EFTs from QCD is presently unknown, they are motivated by QCD and may be verified a posteriori by comparison with experimental data. Among EFTs, the Skyrme model \cite{skyrme} plays a prominent role. In this purely mesonic theory, baryons as well as nuclei emerge as collective, non-perturbative excitations, i.e., topological solitons, where the baryon charge is identified with a topological degree. Restricting to the two-flavor case most relevant for nuclear matter (i.e., to pions as the only field variables), the model originally proposed by Skyrme is given by the following Lagrange density 
\begin{equation}
\mathcal{L}=
\mathcal{L}_2+ \mathcal{L}_4,
\end{equation}
where
\begin{equation}
\mathcal{L}_2=-\lambda_2 \mbox{Tr}\; (L_\mu L^\mu), \;\;\; \mathcal{L}_4=\lambda_4 \mbox{Tr} \; ([L_\mu , L_\nu]^2), \;\;\;  L_\mu \equiv U^\dagger \partial_\mu U
\end{equation}
are the kinetic (sigma model) and Skyrme terms. Further, a potential (non-derivative term) 
\begin{equation}
\tilde{\mathcal{L}}_0(U) = -\tilde \lambda_0 \tilde{\mathcal{U}}(U)
\end{equation}
 is frequently added, e.g., giving masses to the pions. Here, $U$ is the SU(2)-valued Skyrme field. This model is quite successful in the description of light nuclei. Indeed, after the semiclassical quantisation of solitons, the model leads to a remarkably good description of nucleons\cite{nappi}, the deuteron\cite{braaten} and some further light nuclei\cite{carson} (in particular, their spin and isospin excitation spectra\cite{wood,lau}). On the other hand, in the higher baryon charge sector (heavy nuclei) where quite many skyrmions have been calculated numerically\cite{manton}, as well as for infinite nuclear matter, the model suffers from two drawbacks, i.e., the {\it too high binding energies} and the {\it crystalline state of matter}\cite{cryst} for high charge skyrmions, in contrast to basic, qualitative properties of nuclear matter. Recently, several solutions to the binding energy problem have been proposed. One may 
add (infinitely) many vector mesons (the {\it BPS Skyrme vector meson model}\cite{SutBPS,rho1}), include a further, "repulsive" potential (the {\it lightly bound model}\cite{BPSM}), or - this last modification handles both issues at the same time - add a dominant sextic term to the Skyrme lagrangian (the {\it near-BPS Skyrme model}\cite{BPS,Marl1,Marl2,Sp2}).
The resulting lagrangian reads
\begin{equation}  \label{full}
\mathcal{L}=\epsilon\left(\tilde{\mathcal{L}}_0+\mathcal{L}_2+\mathcal{L}_4\right) + \mathcal{L}_{BPS}, 
\end{equation}  
where the BPS part (the {\it BPS Skyrme model})
\begin{equation} \label{BPSmodel}
\mathcal{L}_{BPS} \equiv \mathcal{L}_6+\mathcal{L}_0 \equiv -(24\pi^2)^2 \lambda_6 \mathcal{B}_\mu \mathcal{B}^\mu- \lambda_0 \mathcal{U}
\end{equation}
consists of the sextic term $\mathcal{L}_6$ - the square of the baryon current 
\begin{equation} \label{top-curr}
\mathcal{B}^\mu = \frac{1}{24\pi^2} \epsilon^{\mu \nu \rho \sigma} \mbox{Tr} \; L_\nu L_\rho L_\sigma, \quad B\equiv \int d^3 x \mathcal{B}^0
\end{equation}
($B\ldots$ baryon number) - and a further potential. This new proposal can be regarded as a usual Skyrme model\cite{omega-B,sextic-phen,kop,Ding2007,sextic-skyrme,nitta} (in fact, the most general one if both Poincare symmetry and a standard Hamitonian are assumed) but with a particular choice for the 
parameters. Namely, it is assumed that the parameter $\epsilon$ is chosen such that the standard Skyrme action gives a rather small contribution to the soliton masses, the dominating contribution steming from the BPS term. 
This particular parameter choice is motivated by the very special properties of the BPS part (i.e., the $\epsilon=0$ limit):
\begin{enumerate}
\item {\it BPS nature -- zero binding energies}
\\
Skyrmions in the BPS Skyrme model have energies which linearly depend on the topological (baryon) charge $E_B=C_{\mathcal{U}} |B|$, where the constant $C_{\mathcal{U}}$ depends on the potential. The classical binding energies are, therefore, zero. Finite binding energies may be achieved once semiclassical corrections are taken into account. Further, the model has infinitely many target space symmetries which may be explained by generalized integrability\cite{gen-int}.
\item {\it Perfect fluid }
\\
The BPS Skyrme model describes an effective perfect fluid. Its energy-momentum tensor is of the perfect fluid form, and the static energy functional is {\em SDiff} symmetric\cite{fosco} (invariant under volume-preserving diffeomorphisms (VPDs) on physical space). The model is, in fact, formally completely equivalent to the Eulerian formulation of a relativistic, non-barotropic perfect fluid\cite{chem-pot}.
This result should be contrasted with other field-theoretic models of nuclear matter, which usually do not lead to a perfect fluid at the microscopic (full field theory) level, such that for a perfect-fluid description of nuclear matter a mean-field approximation must be employed. The mean-field limit leads to constant energy and baryon number densities and, therefore, to a barotropic fluid by construction. As we shall see, this difference will be especially important for the description of neutron stars. 
\item {\it Thermodynamics at $T=0$}
\\
As a consequence of the BPS nature of the model, certain bulk observables of static solutions (energy $E$, geometric volume $V$, pressure $P$) are, at the same time, thermodynamical variables obeying the standard thermodynamical relations at zero temperature $T=0$\cite{term}. Further, average densities (average energy density, baryon density, baryon chemical potential) may be easily defined and, again, obey the standard thermodynamical relations. The averaging procedure corresponds to the mean-field theory in other models and gives us the rare occasion to compare exact field theory and mean-field limit results.
\end{enumerate}

These properties are only weakly modified when a small non-BPS part (the usual Skyrme action) is added. Indeed, in the full near-BPS theory small classical binding energies are expected, where the proper shapes of the skyrmions (and their symmetries crucial for the quantisation procedure) should follow from the non-BPS Lagrangian. Furthermore, we get some deviation from the perfect fluid property, which, however, should be subdominant, especially in the high density (pressure) regime.

The features mentioned above are, undoubtedly, physically very well motivated idealisations of nuclear matter and, therefore, allow us to treat the BPS part as the main ingredient of the low energy action. It means that some (but certainly not all) properties of nuclei and nuclear matter are dominated by the BPS Skyrme action and, therefore, one can study them in the BPS sector - at least in a certain approximation. Here enters another remarkable property of the BPS model:
\begin{enumerate}
\item[(4)] {\it Solvability} 
\\
The BPS Skyrme model is a solvable field theory in the sense that one can solve the static field equation exactly for any value of the topological charge, both for the BPS equation (the equilibrium case $P=0$) and for the non-BPS case (non-zero pressure). The solvability leads to an {\it analytical} understanding of some mathematical as well as physical properties of skyrmions and allows to solve the gravitating system with the full back reaction on the matter field taken into account. 
\end{enumerate}

Obviously, although the BPS part of the action provides the most important contribution to the skyrmion masses, it may be treated only as an approximation to the proper EFT, which in this set-up is the near-BPS Skyrme action. The inclusion of the usual (perturbative) Skyrme Lagrangian, corresponding to the inclusion of the dynamical, perturbative pions, which are absent (frozen) in the BPS limit, breaks the {\em SDiff} symmetry and provides shapes for the nuclei. In this sense, the perturbative part of the full near-BPS model is responsible for the surface effects while the BPS part gives the bulk contributions. This leads to an interesting identification of the surface and bulk contributions to observables with the pionic (chiral perturbation theory) and non-perturbative components of the action, respectively. 

To conclude, the near-BPS Skyrme action is a simple purely pionic EFT which provides a unified description of baryons at all scales: from nucleons (where also some properties of the chiral perturbation theory are taken into account) and nuclei to infinite nuclear matter and neutron stars.

We shall frequently use the Skyrme field parametrization
\begin{equation}
U= \exp (i\xi \vec\tau \cdot \vec n) = \cos \xi \, {\bf 1} + i \sin \xi \, \vec \tau \cdot \vec n
\end{equation}
where $\xi$ is a real field, $\vec\tau $ are the Pauli matrices, and
\begin{equation}
\vec n = (\sin \Theta \cos \Phi ,\sin \Theta \sin \Phi ,\cos \Theta )
\end{equation}
is a three-component unit vector. Our metric convention is $(+,-,-,-)$. Further, we sometimes use the axially symmetric ansatz (here $(r,\theta ,\phi)$ are either spherical polar coordinates or Schwarzschild coordinates, respectively)
\begin{equation} \label{ax-sym}
\xi = \xi (r) \, , \quad \Theta = \theta \, , \quad \Phi = B\phi
\end{equation}
leading to a spherically symmetric baryon density $\mathcal{B}^0 = \mathcal{B}^0 (r)$ and - for the BPS submodel - to a spherically symmetric energy density. For the full near-BPS model, the axially symmetric ansatz is incompatible with the field equations (except for the hedgehog case $B=1$).

All potentials $\mathcal{U}$ we consider depend on $U$ only via its trace, i.e., only depend on $\xi$, such that, while chiral symmetry is broken, isospin remains a symmetry. Further, their unique vacuum is at $U={\bf 1}$, i.e., at $\xi = 0$ (i.e., $\mathcal{U}=\mathcal{U}(\xi)$, and $\mathcal{U}(\xi =0)=0$). 

As a matter of fact, two different potentials show up in the full near-BPS Skyrme model (\ref{full}), namely the potential $\tilde{\mathcal{U}}$ of the "perturbative" part and the potential $\mathcal{U}$ of the BPS part. As the perturbative potential $\tilde{\mathcal{U}}$ comes together with the Dirichlet (non-linear sigma model) term $\mathcal{L}_2$, it is natural to set it equal to the pion mass potential 
\begin{equation}
\tilde{\mathcal{U}} = \mathcal{U}_\pi \equiv (1/2) {\rm tr}(1-U)=1-\cos \xi,
\end{equation} 
such that the resulting mass parameter may be related to the physical pion mass. As a consequence, the BPS potential $\mathcal{U}$ should not contain a mass-like contribution (i.e., a term quadratic in the pion field). A simple and natural assumption is that the BPS potential is quartic in the pion field close to the vacuum, the simplest choice being just $\mathcal{U} = \mathcal{U}_\pi^2$, i.e., the square of the pion mass potential. We shall specify our choices for $\mathcal{U}$ when required. There is a second reason why the BPS potential should contain no quadratic contribution, namely the low regularity of the resulting static (compact) BPS skyrmion solutions\cite{Marl1}. These solutions are continuous but not differentiable at the compacton boundary (soliton surface). This low regularity does not cause problems in the pure BPS Skyrme model (there the physical quantities calculable from the soliton solutions remain finite and well-defined) but becomes problematic once one wants to go beyond the BPS submodel in a perturbative fashion. The Dirichlet (non-linear sigma model) energy $E_2 = -\int d^3 x \mathcal{L}_2$, e.g., becomes infinite when evaluated for such low regularity BPS skyrmion solutions. 
 
\section{The BPS Skyrme model}\label{bps}

The BPS limit of the near BPS action is obtained by taking $\epsilon=0$ (here, $\lambda^2 \equiv (24)^2 \lambda_6$, $\mu^2 \equiv \lambda_0$)
\begin{equation}
\mathcal{L}_{BPS} =-\lambda^2\pi^4 \mathcal{B}_\mu \mathcal{B}^\mu- \mu^2 \mathcal{U}.
\end{equation}
This limit is nontrivial both from a physical and a mathematical point of view. First of all, if $\epsilon=0$ there is no kinetic term for the Skyrme field and, consequently, there are no perturbative pions. There are still pionic degrees of freedom $\vec{\pi}$ defined as usually by the Skyrme field, $\vec \pi = \sin \xi \, \vec n$, but they do not propagate since there is no $(\partial \vec{\pi})^2$ term. One may say that in this limit perturbative excitations are suppressed while coherent (topological) excitations survive.  

From a mathematical perspective, we deal with a non-analytical limit (perturbation), as for a non-zero value of $\epsilon$ the perturbative part always dominates near the vacuum. Moreover, for $\epsilon=0$ one has an enormous enhancement of symmetry, which changes from a finite-dimensional group to the infinite-dimensional VPD group. 

Let us notice that the model is based on terms which are related to collective, nonperturbative properties of strong interactions: chiral symmetry breaking for the potential and Skyrme field topology for the sextic term. Therefore, it might be expected to be relevant whenever nonperturbative properties should be important like, for instance, in regions of not too small baryon density (as is the case, e.g., inside nuclei or neutron stars).

\subsection{Bogomolny bound and BPS equation}

The first crucial property of the BPS Skyrme model is that the energy functional for static configurations
\begin{equation}
E=\int d^3 x \left( \pi^4\lambda^2 \mathcal{B}_0^2+\mu^2 \mathcal{U} \right)
\end{equation}
has a Bogomolny bound and infinitely many BPS solutions saturating the bound. \cite{BPS,Sp1} To derive the bound, it is useful to recall that the target space $SU(2)$ as a manifold is just the three-sphere $\mathbb{S}^3$ and the topological charge density three-form $\mathcal{B}_0d^3x \equiv \mathcal{B}_0 {\rm vol}_{\mathbb{R}^3} $ is proportional to the pullback (under the map $U: \mathbb{R}^3 \rightarrow \mathbb{S}^3$) of the volume form ${\rm vol}_{\mathbb{S}^3}$ on $\mathbb{S}^3$, i.e., 
\begin{equation} \label{vol-form}
\mathcal{B}_0 \,  {\rm vol}_{\mathbb{R}^3} = \pm \frac{1}{2\pi^2} U^*({\rm vol}_{\mathbb{S}^3}).
\end{equation} 
Then, the bound is
\begin{eqnarray}
E&=& \int d^3 x \left( \pi^2 \lambda \mathcal{B}_0 \pm \mu \sqrt{\mathcal{U}} \right)^2 \mp 2 \pi^2 \lambda \mu \int d^3x   \mathcal{B}_0  \sqrt{\mathcal{U}} \\
&\geq& 2 \pi^2 \lambda \mu \left|\int d^3x   \mathcal{B}_0  \sqrt{\mathcal{U}} \right| = 2\pi^2 \lambda \mu |B| \left< \sqrt{\mathcal{U}}\right>_{\mathbb{S}^3}
\end{eqnarray}
where 
\begin{equation}
 \left< \sqrt{\mathcal{U}}\right>_{\mathbb{S}^3} \equiv \frac{1}{2\pi^2} \int_{\mathbb{S}^3}  {\rm vol}_{\mathbb{S}^3}\sqrt{\mathcal{U}}
\end{equation}
is the average value of $\sqrt{\mathcal{U}}$ on the target space. For potentials $\mathcal{U}(\xi)$ we may use ${\rm vol}_{\mathbb{S}^3} = \sin^2 \xi \sin \Theta d\xi d\Theta d\Phi$ and the average simplifies to 
\begin{equation} \label{simp-average}
 \left< \sqrt{\mathcal{U}}\right>_{\mathbb{S}^3} = \frac{2}{\pi} \int_0^\pi d\xi \sin^2 \xi \sqrt{\mathcal{U}(\xi)}.
\end{equation}
The above bound is saturated by solutions of the BPS equation
\begin{equation} \label{BPS-eq}
\pi^2 \lambda \mathcal{B}_0 \pm \mu \sqrt{\mathcal{U}}=0.
\end{equation}
Owing to the {\em SDiff} symmetry, there exist infinitely many solutions with arbitrary shapes for each $B$.

\subsection{Perfect fluid}
The BPS Skyrme model has the symmetries \cite{fosco} (the {\em SDiff} symmetries on physical space) and the energy-momentum tensor of a perfect fluid.
For the derivation of the energy-momentum tensor it is useful to introduce a nontrivial metric $g_{\rho \sigma}$ (here $g= {\rm det} g_{\rho\sigma}$), then 
the action of the BPS Skyrme model is
\begin{equation}
S_{06} = \int d^4 x |g|^\frac{1}{2} \left( -\lambda^2 \pi^4 |g|^{-1} g_{\rho\sigma} {\cal B}^\rho {\cal B}^\sigma - \mu^2 {\cal U} \right) ,
\end{equation}
and the corresponding  energy-momentum tensor is
\begin{equation}
T^{\rho\sigma} = -2|g|^{-\frac{1}{2}}\frac{\delta}{\delta g_{\rho\sigma}} S_{06} = 2\lambda^2 \pi^4 |g|^{-1} {\cal B}^\rho {\cal B}^\sigma - \left( \lambda^2 \pi^4 |g|^{-1} g_{\pi\omega} {\cal B}^\pi {\cal B}^\omega - \mu^2 {\cal U} \right) g^{\rho\sigma} .
\end{equation}
Here, ${\cal B}^\mu$ is still as defined in (\ref{top-curr}), i.e., in a non-flat space-time it is a first rank tensor density rather than a vector. We could introduce the contravariant vector $\tilde{\cal B}^\mu = |g|^{-\frac{1}{2}}{\cal B}^\mu$ but prefer to use ${\cal B}^\mu$ for simplicity. 
The above energy-momentum tensor is, in fact, the energy-momentum tensor of a perfect fluid,
\begin{equation} \label{perf-fluid}
T^{\rho \sigma} = (p+\varepsilon )u^\rho u^\sigma - pg^{\rho\sigma}
\end{equation}
where the four-velocity $u^\rho$, energy density $\varepsilon$ and pressure $p$ are
\begin{equation}
u^\rho = {\cal B}^\rho / \sqrt{g_{\sigma \pi} {\cal B}^\sigma {\cal B}^\pi}
\end{equation}
\begin{eqnarray}
\varepsilon &=& \lambda^2 \pi^4 |g|^{-1} g_{\rho\sigma} {\cal B}^\rho {\cal B}^\sigma + \mu^2 {\cal U} \nonumber \\
p &=& \lambda^2 \pi^4 |g|^{-1} g_{\rho\sigma} {\cal B}^\rho {\cal B}^\sigma - \mu^2 {\cal U} .
\end{eqnarray}
In Minkowski space and for a static field configuration in cartesian coordinates, the non-zero components of the energy-momentum tensor are
\begin{equation} \label{varepsilon}
T^{00} = \varepsilon \equiv \lambda^2 \pi^4 \mathcal{B}_0^2 + \mu^2 \mathcal{U} ,
\end{equation}
\begin{equation} \label{p}
T^{ij} = \delta^{ij} p  \equiv \delta^{ij} \left(\lambda^2 \pi^4 \mathcal{B}_0^2 - \mu^2 \mathcal{U} \right) .
\end{equation}
The analogy with a perfect fluid goes, in fact, much further. If the three variables of the Skyrme field (e.g. $\xi$, $\Theta$ and $\Phi$) are formally identified with the co-moving coordinates of a perfect fluid, then the BPS Skyrme model is formally completely equivalent to a non-barotropic, relativistic perfect fluid in the Eulerian formulation\cite{chem-pot}.

\section{Thermodynamics at $T=0$}

\subsection{Equation of state}
Energy-momentum conservation implies that the pressure of static solutions must be constant, 
\begin{equation}
\partial_i T^{ij} = \delta^{ij} \partial_i p =0 \quad\Rightarrow \quad p =P=\mbox{const}.
\end{equation}
It may, in fact, be demonstrated that the constant pressure equation (here $P$ is a non-negative constant)
\begin{equation}
 p \equiv  \lambda^2 \pi^4 \mathcal{B}_0^2 - \mu^2 \mathcal{U} =P
\end{equation}
is a first integral of the static field equations, where $P$ is the corresponding integration constant\cite{term}. It may be re-written like
\begin{equation} \label{const-pres-2}
\mathcal{B}_0 = \pm \frac{\mu}{\lambda \pi^2} \sqrt{\mathcal{U} + \tilde P} \, , \quad \tilde P \equiv \frac{P}{\mu^2}
\end{equation}
and, obviously, generalizes the BPS equation (\ref{BPS-eq}) to nonzero pressure (we shall assume the plus sign and $B>0$ - baryons, not antibaryons - in the sequel). Its first-order nature allows to derive the thermodynamics of the model at zero temperature $T=0$. 
Indeed, from (\ref{vol-form}) and the above constant-pressure equation we find for the volume form
\begin{equation}
{\rm vol}_{\mathbb{R}^3} = \frac{1}{2\pi^2 \mathcal{B}_0} U^* \left( {\rm vol}_{\mathbb{S}^3} \right) = \frac{1}{2}\frac{\lambda}{\mu} 
U^* \left( \frac{{\rm vol}_{\mathbb{S}^3}}{\sqrt{ \mathcal{U} + \tilde P}} \right)
\end{equation}
and for the geometric volume
\begin{equation} \label{volume}
V(P) = \int_\Omega {\rm vol}_{\mathbb{R}^3} = \frac{B}{2}\frac{\lambda}{\mu} \int_{\mathbb{S}^3} 
 \frac{{\rm vol}_{\mathbb{S}^3}}{\sqrt{ \mathcal{U} + \tilde P}} = 
\pi^2 B\frac{\lambda}{\mu} \left< \frac{1}{\sqrt{ \mathcal{U} + \tilde P}} \right>_{\mathbb{S}^3}. 
\end{equation}
Here, $\Omega \subset {\mathbb{R}^3}$ is the locus set of the skyrmion, i.e., the set of points where the static solution $U(\vec r)$ deviates from its vacuum value. It follows that all static skyrmions with the same pressure have the same volume. For nonzero pressure, all skyrmions have finite volume. For zero pressure (BPS skyrmions or equilibrium solutions) it follows easily from Eq. (\ref{simp-average}) that the geometric volume is finite for potentials with a near-vacuum behaviour like $\lim_{\xi \to 0 } \mathcal{U}(\xi) \sim \xi^\alpha$ for $\alpha <6$. These BPS skyrmions are, therefore, compactons.\cite{term}

In a similar fashion, we find for the on-shell energy density
\begin{equation}
\varepsilon = \lambda^2 \pi^4 \mathcal{B}_0^2 + \mu^2 \mathcal{U} = 2\mu^2 \mathcal{U} + P
\end{equation}
and for the energy
\begin{equation}
E(P)= \int_\Omega {\rm vol}_{\mathbb{R}^3} (2\mu^2 \mathcal{U} + P ) = 
\pi^2 B\lambda \mu \left<  \frac{2\mathcal{U} + \tilde P}{\sqrt{ \mathcal{U} + \tilde P}} \right>_{\mathbb{S}^3}. 
\end{equation}
So, both energy $E$ and volume $V$ take the same value for all static solutions with the same value of the pressure $P$. In addition, it may be shown easily that $V$ and $E$ obey the standard thermodynamical relation
\begin{equation}
P = -\left( \frac{\partial E}{\partial V} \right)_B .
\end{equation}
In other words, the bulk observables $P$, $V$ and $E$ are, at the same time, standard thermodynamcial variables, although they were introduced in a purely field-theoretic context ($P$ as an integration constant, $V$ as the geometric volume, and $E$ as the field energy of a static skyrmion). Consequently, Eq. (\ref{volume}) is the (global) equation of state (EoS) (in terms of bulk thermodynamic variables) of our system. Particular EoS result from particular choices for the potential $\mathcal{U}(\xi)$.

From the above results it is easy to calculate further thermodynamical variables. One quantity of special interest in nuclear physics is the compression modulus
\begin{equation}
\mathcal{K} = \frac{9V^2}{B}\left( \frac{\partial^2 E}{\partial V^2} \right)_B
\end{equation}
which is related to the (isothermal; but remember that in our case $T=0$) compressibility $\kappa$,
\begin{equation}
\mathcal{K} = \frac{9V}{B\kappa }  \;  , \qquad 
\kappa \equiv -\frac{1}{V}\left( \frac{\partial V}{\partial P} \right)_B .
\end{equation}
The compression modulus of nuclear matter at nuclear saturation density (i.e., at equilibrium $P=0$) is known to be about $\mathcal{K}(P=0) \simeq 250\, \mbox{MeV}$. This seems to cause a problem for Skyrme models. Indeed, if one (inappropriately) assumes a constant baryon density (i.e., assumes a mean-field limit) for skyrmionic matter, such that the softest volume-changing excitation is the uniform (Derrick) rescaling $\vec r \to \Lambda \vec r$, then the resulting compression modulus is much larger than its physical value\cite{term}.  Using the thermodynamical results just derived, it turns out that the compression modulus in the BPS Skyrme model is, in fact, zero for realistic potentials. One easily calculates
\begin{equation}
\kappa (P=0) = \frac{\pi B \lambda}{\mu^3 V(0)} \int_0^\pi d\xi \sin^2 \xi \mathcal{U}^{-\frac{3}{2}}
\end{equation}
which is infinite for potentials $\lim_{\xi \to 0}\mathcal{U} \sim \xi^\alpha$ for $\alpha \ge 2$, corresponding to a zero compression modulus. 
It is expected that a more complete treatment (where both the additional terms of the near-BPS Skyrme model and the collective coordinate quantization are taken into account) may lead to a compression modulus which is closer to its physical value. 

Only for the limiting case of a constant (step function) potential $\mathcal{U}_\Theta \equiv \Theta (\xi)$ (leading to a constant baryon density, see Eq. (\ref{const-pres-2})) the mean-field argument is correct, and the compression modulus results too high already in the BPS Skyrme model. This potential is, however, not realistic. 

For potentials which lead to compact BPS skyrmions (i.e., for $\alpha <6$), a liquid-gas phase transition occurs in the model. Indeed, in this case static charge $B$ solutions in a gaseous phase at zero pressure with $V>V_0 \equiv V(P=0)$ exist, which are just collections of non-overlapping compactons of smaller (e.g. $B=1$) charges, where the additional volume $\delta V = V-V_0$ is occupied by the empty space (vacuum) surrounding the compactons. At $V=V_0$, a phase transition to a liquid phase described by the EoS (\ref{volume}) occurs. Interestingly, this is precisely equivalent to the liquid-gas phase transition of nuclear matter at nuclear saturation, so the model exactly reproduces the conjectured phase diagram of QCD at zero temperature close to nuclear saturation\cite{fuku-hatsu}.

\subsection{Local densities}
The dynamical and thermodynamical properties of a (perfect) fluid are usually described in terms of some densities and their relations. In our case, these are the energy density $\varepsilon$, the pressure $p$ and the particle (baryon) number density $\rho_B \equiv \mathcal{B}_0$. Off-shell (without using the static field equations) each density is a certain function of the Skyrme field and its first derivative, see Eqs. (\ref{top-curr}), (\ref{varepsilon}) and (\ref{p}). They are related by the universal (off-shell and potential-independent) relation
\begin{equation}
\varepsilon + p = 2\lambda^2 \pi^4 \rho_B^2 .
\end{equation}
Further, we may introduce the (local) baryon chemical potential $\boldsymbol{\mu}$ via the well-known relation
\begin{equation} \label{loc-chempot-def}
\varepsilon + p =  \rho_B \boldsymbol{\mu}
\end{equation}
which leads to the universal off-shell relation
\begin{equation}
 \boldsymbol{\mu} = 2\pi^4 \lambda^2 \rho_B,
\end{equation}
 i.e., the local baryon chemical potential is exactly proportional to the baryon density. $\varepsilon$ and $p$ are related by the local off-shell energy-density--pressure EoS
\begin{equation}
\varepsilon = p + 2\mu^2 \mathcal{U}
\end{equation}
where $\mathcal{U}$ is a function of the Skyrme field. On-shell the pressure is constant, $p=P$, whereas the remainig densities are non-constant and solution-dependent,
\begin{eqnarray} \label{on-shell-eos}
\varepsilon = \varepsilon (P,\vec r) &\equiv& P+ 2\mu^2 \mathcal{U}(P,\vec r) \\
\rho_B = \rho_B (P,\vec r) &\equiv& \frac{1}{\lambda \pi^2} \sqrt{ P+ \mu^2 \mathcal{U}(P,\vec r)}
\end{eqnarray}
(and, of course, $\boldsymbol{\mu} (P, \vec r)= 2\pi^4 \lambda^2 \rho_B (P, \vec r)$). In particular, the local EoS (\ref{on-shell-eos}) explicitly depends on $\vec r$, so $\varepsilon$ and $P$ are {\em not} related by an algebraic relation. The perfect fluid described by the BPS Skyrme model is, therefore, {\em non-barotropic} (except for the special case of the step function potential). On the other hand, the three remaining densities $\varepsilon$, $\rho_B$ and $\boldsymbol{\mu}$ are related algebraically on-shell. 
 
\subsection{Mean-field limit}
Owing to the perfect-fluid form and the BPS property of the BPS Skyrme model, thermodynamical variables and densities may be found exactly directly from the underlying field theory, without the necessity of any further approximation, like a thermodynamical or mean-field (MF) limit. This distinguishes the BPS Skyrme model from other models of nuclear matter, where usually a MF limit is performed to arrive at a perfect fluid allowing for a thermodynamical description. In such a MF limit, the resulting densities are constant, by construction, and the corresponding perfect fluid is, therefore, barotropic. For a direct comparison of BPS Skyrme model results with the results of other models, a MF limit of the BPS skyrmion thermodynamics would be helpful, and we shall see that such a limit may be easily performed.  In addition to facilitating comparisons, this limit provides us with the unique opportunity to confront exact and MF theory results within the same field theory. 

Indeed, the MF energy density and baryon density are just the on-shell volume averages of the on-shell energy and baryon number,
\begin{eqnarray} \label{MFEoS}
\bar \varepsilon &\equiv& \frac{E(P)}{V(P)} = \mu^2 \frac{\left<  \frac{2\mathcal{U} + \tilde P}{\sqrt{ \mathcal{U} + \tilde P}} \right>_{\mathbb{S}^3}}{
\left< \frac{1}{\sqrt{ \mathcal{U} + \tilde P}} \right>_{\mathbb{S}^3}} \\
\label{bar-rho}
\bar \rho_B &\equiv& \frac{B}{V(P)} = \frac{\mu}{\pi^2 \lambda} \frac{1}{
\left< \frac{1}{\sqrt{ \mathcal{U} + \tilde P}} \right>_{\mathbb{S}^3}} 
\end{eqnarray}
(obviously, the average pressure $\bar p \equiv V^{-1} \int d^3 x P =P$ coincides with its constant on-shell value $P$). In particular, the energy-density--pressure EoS 
(\ref{MFEoS}) is now barotropic, $\bar \varepsilon = \bar \varepsilon (P)$, as is generally the case within MF theory. The expression for $\bar \varepsilon (P)$ allows to calculate the MF speed of sound $\bar v_s$ via $\bar v_s^{-2} = (\partial \bar \varepsilon / \partial P)_B$.

There are two possible definitions for the MF chemical potential $\bar{\boldsymbol{\mu}} $. The first one is just the MF version of Eq. (\ref{loc-chempot-def}), i.e., 
\begin{equation} \label{MF-chempot-def}
\bar\varepsilon + P =  \bar\rho_B \bar{\boldsymbol{\mu}} .
\end{equation}
Upon integrating Eq. (\ref{loc-chempot-def}) and comparing with (\ref{MF-chempot-def}), it follows that
\begin{equation}
\bar{\boldsymbol{\mu}} = \frac{1}{B} \int d^3 x \rho_B\boldsymbol{\mu} 
\end{equation}
so $\bar{\boldsymbol{\mu}} $ is defined as a baryon number average, not as a volume average. 
The second, well-known definition is
\begin{equation}
\bar{\boldsymbol{\mu}} =\left( \frac{\partial E}{\partial B}\right)_V
\end{equation}
i.e., the change in energy by adding a particle at constant volume. Obviously, consistency of our thermodynamcial description requires that the two definitions of $\bar{\boldsymbol{\mu}} $ coincide. From the constant volume condition $((\partial V)/(\partial B)) =0$, the relation
\begin{equation}
\frac{B}{2} \left< \frac{1}{(\mathcal{U} + \tilde P)^\frac{3}{2} } \right>_{\mathbb{S}^3} \left( \frac{\partial \tilde P}{\partial B} \right)_V = 
\left< \frac{1}{(\mathcal{U} + \tilde P)^\frac{1}{2} } \right>_{\mathbb{S}^3}
\end{equation}
follows, and with its help, indeed, one easily proves that both definitions lead to the same target space average 
\begin{equation} \label{MF-chempot}
\bar{\boldsymbol{\mu}} = \left( \frac{\partial E}{\partial B}\right)_V = \frac{E+PV}{B} = 2\pi^2 \lambda \mu \left< \sqrt{\mathcal{U} + \tilde P} \right>_{\mathbb{S}^3}.
\end{equation} 
We remark that the MF chemical potential is, in general, {\em not} linear in $\bar \rho_B$, i.e., $\bar{\boldsymbol{\mu}}  \not= 2\pi^4 \lambda^2 \bar \rho_B$. We further remark that the above equation (\ref{MF-chempot}) is valid both for the liquid ($\rho_B \ge \rho_{B,0} \equiv (B/V_0)$) and for the gaseous phase ($\rho_B < \rho_{B,0}$). Indeed, in the gaseous phase $P=0$, so the MF chemical potential takes the constant value $\bar{\boldsymbol{\mu}} = (E/B) = E_{1,0} \equiv E(B=1,P=0)$, i.e., the energy it costs to add one $B=1$ BPS skyrmion to the gas of non-overlapping compact BPS skyrmions.

Another quantity of physical interest is the energy per baryon number
\begin{equation}
\bar\varepsilon_B \equiv \frac{E}{B} = \bar{\boldsymbol{\mu}} - \frac{P}{\bar \rho_B},
\end{equation}
which may be interpreted as the "in-medium" energy of a $B=1$ skyrmion, i.e., the energy of a $B=1$ skyrmion in the environment of skyrmionic matter. Obviously, the corresponding in-medium volume of a $B=1$ skyrmion is just ${\mathcal{V}}_B = \bar\rho_B^{-1}$.

\subsection{An example}
\begin{figure}[t]
\begin{center}
\includegraphics[width=.7\textwidth]{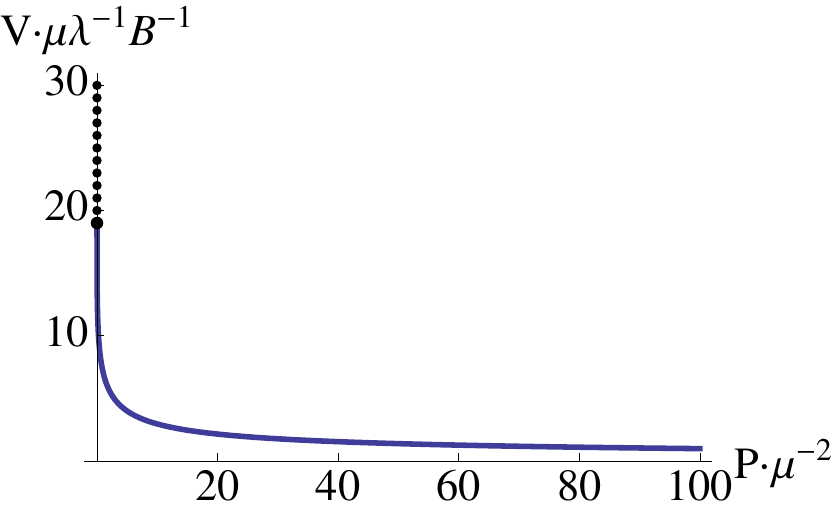}
\caption{The PV diagram for the quartic potential $\mathcal{U} = \mathcal{U}_\pi^2$, in rescaled (dimensionless) units. The dots correspond to the gaseous phase, whereas the continuous line describes the liquid phase.}
\label{PV-diag}
\end{center}
\end{figure}

\begin{figure}[t]
\begin{center}
\includegraphics[width=.7\textwidth]{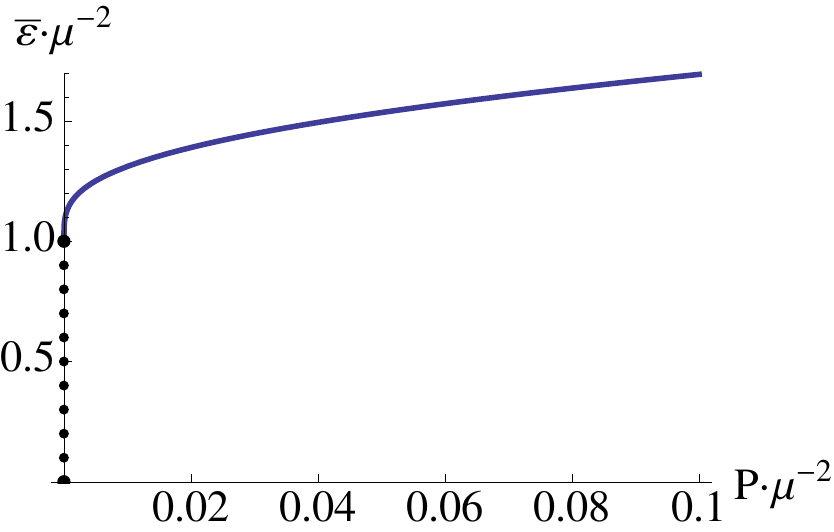}
\caption{The MF energy density as a function of the pressure, both in rescaled (dimensionless) units. The dots correspond to the gaseous phase. It can be seen that the high-density behaviour $\bar \varepsilon \propto P$ sets in rather soon.}
\label{ebar-p}
\end{center}
\end{figure}

\begin{figure}[t]
\hspace*{-0.5cm}\includegraphics[width=.5\textwidth]{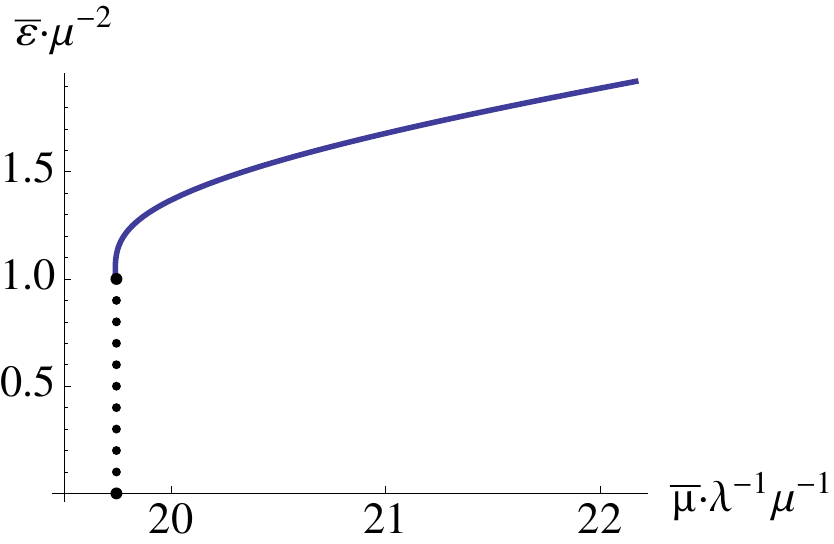}
\hspace{0.2cm}
\includegraphics[width=.5\textwidth]{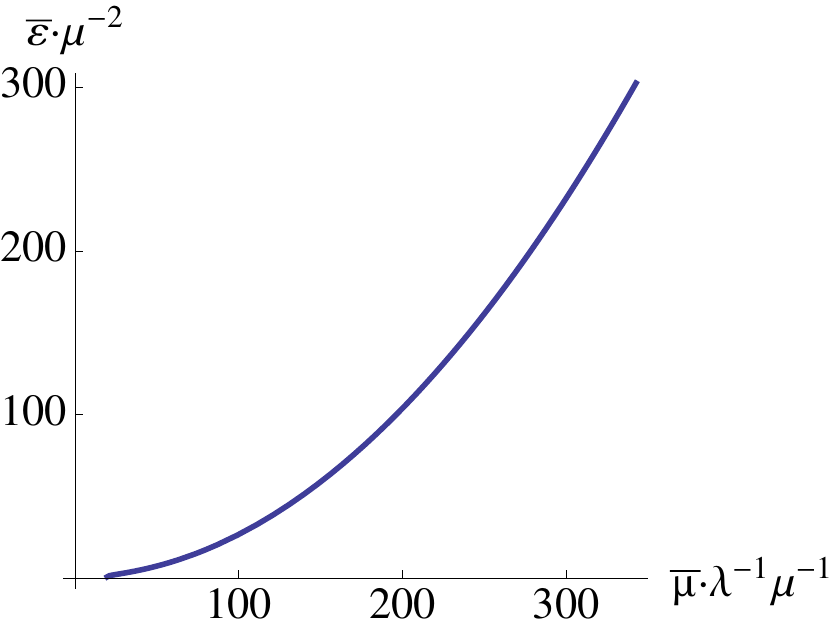} 
\caption{The MF energy density as a function of the MF baryon chemical potential. The dots correspond to the gaseous phase. Low-density region (left) and high-density region (right).}
\label{ebar-mubar}
\end{figure}

Here we want to study in some more detail the thermodynamical relations of a particular example, namely the pion mass potential squared, $\mathcal{U} = \mathcal{U}_\pi^2$. In this case, the explicit expressions for the energy and volume are 
\begin{equation}
E= B\lambda \mu \cdot \pi^2 \frac{1}{\sqrt{ \frac{P}{\mu^2} } } \left( \frac{P}{\mu^2} \; _3F_2 \left[  \left\{ \frac{1}{2}, \frac{3}{4}, \frac{5}{4}\right\}, 
\left\{ \frac{3}{2},2\right\}, -\frac{4\mu^2}{P} \right] + \right. \nonumber
\end{equation}
\begin{equation}
\left. + \frac{5}{2}  \; _3F_2 \left[  \left\{ \frac{1}{2}, \frac{7}{4}, \frac{9}{4}\right\}, 
\left\{ \frac{5}{2},3\right\}, -\frac{4\mu^2}{P} \right] \right) ,
\end{equation}
\begin{equation}
V= \frac{B \lambda}{\mu}\cdot  \pi^2 \frac{1}{\sqrt{ \frac{P}{\mu^2} } } \; _3F_2 \left[  \left\{ \frac{1}{2}, \frac{3}{4}, \frac{5}{4}\right\}, 
\left\{ \frac{3}{2},2\right\}, -\frac{4\mu^2}{P} \right] ,
\end{equation}
(here ${}_pF_q[\{ a_1,\ldots ,a_p\} ,\{ b_1, \ldots ,b_q\} ,z]$ is a generalized hypergeometric function) from which the MF thermodynamic densities may be calculated. It is, however, more instructive to plot the resulting phase diagrams. 
In Fig. \ref{PV-diag} we plot the P-V diagram. Both the liquid-gas phase transition and the leading high-density behaviour $P\sim V^{-2}$ are clearly visible. In Fig. \ref{ebar-p} we plot the MF energy density as a function of the pressure, i.e., the EoS $\bar \varepsilon (P)$. This EoS is soft very close to $P=0$, but rather soon approaches the maximally stiff limit $\bar \varepsilon = P + \, $const. Indeed, the speed of sound is zero at $P=0$ but quickly approaches the maximum value $\bar v_s =1$ for nonzero $P$. The behaviour close to $P=0$ is, in fact, determined just by the (in this case, quartic) vacuum approach of the potential.
 In Fig. \ref{ebar-mubar} we plot the dependence of $\bar \varepsilon$ on the MF baryon chemical potential. In the high-density limit, they approach the universal, potential-independent relation $\bar \varepsilon \propto \bar{\boldsymbol{\mu}}^2$.

\subsection{Some implications}

The importance of our thermodynamical results for the description of nuclear matter was discussed in detail in Refs. \refcite{term,chem-pot,eos}. Here we just want to emphasize the most important points.

\begin{itemize}
\item 
Our system for sufficiently large $B$ (and for general pressure) may be considered a model for infinite nuclear matter, at least in a first approximation. Infinite nuclear matter is an idealized system of nuclear matter, where surface contributions, Coulomb energy contributions and the difference between protons and neutrons are ignored, such that effectively only the effects of the strong interaction are present. In our model, surface effects are absent owing to the BPS property, the Coulomb interaction is not taken into account (although it can be included, see next  section), and the same classical $B=1$ skyrmion describes both a proton and a neutron, i.e., a nucleon. The static energy of infinite nuclear matter per nucleon at equilibrium (at saturation density) $E_{\rm nm}$ should therefore be identified with our energy per baryon $E_{1,0}\equiv E_{B=1,P=0}$ at zero pressure, and the density of infinite nuclear matter at equilibrium (the nuclear saturation density $\rho_s$) with our baryon density at zero pressure. We shall use the nuclear physics values
\begin{equation} \label{nuc-sat-values}
E_{\rm nm}= E_{1,0} =    923.3 \; \mbox{MeV} \; , \quad \rho_s = \bar\rho_{B,0} = 0.153 \; \mbox{fm}^{-3},
\end{equation}
where $E_{\rm nm} = E_{\rm N}-E_{\rm b}$, $E_{\rm N} = 939.6 \; \mbox{MeV}$ is the nucleon mass, and $E_{\rm b} = 16.3 \; \mbox{MeV}$ is the binding energy per nucleon of infinite nuclear matter. These values may be used to obtain fit values for our coupling constants $\lambda$ and $\mu$, where the concrete fit values depend on the potential (see section 5). 
\item
Infinite nuclear matter at saturation density is an equilibrium configuration of nuclear matter at zero pressure but finite density $\rho_s$, where the equilibrium is the result of a precise balance between attractive and repulsive forces. In relativistic field theory models of nuclear matter (Walecka model, Quantum Hadron Dynamics (QHD), etc.) this results from a balance between the repulsive force induced by the omega meson and the attractive force due to the sigma meson (and possibly further mesons), where the exact balance has to be achieved by a fine tuning of the coupling constants. In the BPS Skyrme model, this balance is an automatic consequence of the BPS property and of the exactly inverse Derrick scaling of the sextic ($\mathcal{L}_6$) and the potential ($\mathcal{L}_0$) terms. Here, $\mathcal{L}_6$ tends to expand the field configuration, corresponding to a repulsive force, whereas $\mathcal{L}_0$ tends to collapse it (corresponding to an attractive force). It is, therefore, natural to relate the sextic term to the omega meson term of other EFTs, $\mathcal{L}_6 \sim \mathcal{L}_\omega$, and the potential to the sigma meson, $\mathcal{L}_0 \sim\mathcal{L}_\sigma$. This identification is further strengthened by the facts that the sextic term $\mathcal{L}_6 $ is the leading contribution in a derivative expansion of the $\omega$ meson term in a vector-meson extended Skyrme model, and that both the potential and the sigma meson are related to chiral symmetry breaking. 
\item
For all EFTs of nuclear matter containing either the $\omega$ meson (e.g., Walecka model, or QHD) or a fermionic vector-vector interaction (of quark currents, like the Nambu--Jona-Lasinio (NJL) model or the vector-enhanced bag model), the repulsion due to the $\omega$ meson or the vector-vector interaction dominates in the limit of high density, leading to an asymptotic equation of state $\bar \varepsilon \sim P + \ldots$. But this is precisely the asympotic EoS of the BPS Skyrme model, and it is the sextic term which is responsible for this behaviour. Indeed, in the limit of $p>> \mu^2 \mathcal{U}$ we get $\varepsilon = p = \lambda^2 \pi^4 \rho_B^2$. For the MF energy density we may easily find the next-to-leading term,
\begin{equation}
\bar \varepsilon = P + B_\infty \; , \quad B_\infty \equiv 2\mu^2 \left< \mathcal{U} \right>_{\mathbb{S}^3},
\end{equation}
where $B_\infty$ is a kind of asymptotic bag constant. This asymptotic agreement between the EoS even allows to quantitatively relate a parameter of the BPS Skyrme model with $\omega$ meson parameters\cite{chem-pot}, concretely $\pi^4 \lambda^2 = (1/2)(g_\omega^2/m_\omega^2)$ (here $m_\omega$ and $g_\omega$ are the mass and coupling constant of the $\omega$ meson). 
\end{itemize}
We conclude that the inclusion of the sextic term seems to be mandatory
for a realistic description of infinite nuclear matter in a Skyrme model context, both at nuclear saturation and in the limit of high density.
Further, the physical effects of the sextic term are equivalent to the physical effects of the $\omega$ meson in other EFTs of nuclear matter, so we may say that the $\omega$ meson in the (near) BPS Skyrme model is "hidden" in the specific choice of the action (the sextic term) and in the resulting particular behaviour of its solutions. In other words, although the $\omega$ meson is not introduced as an independent field variable, it reappears as an emergent object (a collective excitation), similarly to the baryons themselves (which appear as topological solitons). 

\section{Nuclear binding energies}\label{be}
One important application of the near-BPS Skyrme model is the calculation of nuclear binding energies, given that the original Skyrme model leads to too large binding energies already classically, and that one strong motivation for the near-BPS model is the possibility to reduce them. The most accurate way to proceed seems to consist in numerically calculating near-BPS skyrmions for different baryon numbers and, further, calculate some corrections to the classical soliton energies (rigid rotor quantization of spin and isospin, Coulomb energy corrections), analogously to what is done for the standard Skyrme model. The (small number of) coupling constants of the model may then be fixed by fitting to a small number of nuclear data and, once this is done, the calculated energies represent the model predictions for nuclear binding energies which may be compared with experimental results. Here, the BPS part of (minus) the action (the energy functional in the static case) will provide the main contributions to the classical soliton energies. On the other hand, as this BPS part gives exactly the same energy for skyrmions of arbitrary shapes (due to its SDiff invariance), it is not unreasonable to assume that the skyrmion shapes will essentially be determined by the perturbative (standard Skyrme) part, such that the successes of the standard Skyrme model in the prediction of spin/isospin excitational spectra may be maintained. Unfortunately, the reliable numerical calculation of near-BPS skyrmions is a difficult task which is beyond the scope of the present contribution. Some first results on the numerical calculation of near-BPS skyrmions are presented in Ref. \refcite{BPSM}, where it is found that in the near-BPS limit (i.e., for sufficiently small $\epsilon$), the numerical calculations are hampered by large gradients which are related to the appearence of small regions of zero baryon density. Although certainly rewarding, a full numerical analysis of the near-BPS Skyrme model, therefore, requires a significant refinement of the numerical tools employed up to now.  

Barring full numerical calculations, the next simpler option seems to be a perturbative approach, where a solution of the BPS submodel is inserted into the "perturbative" part of the near-BPS model for an approximate determination of the corresponding skyrmion energies. This approach is complicated by the fact that the BPS submodel has infinitely many solutions with arbitrary shapes, so that the "perturbative" energy should be minimized over these solutions, in the spirit of degenerate perturbation theory. The problem of minimizing the Dirichlet and Skyrme energy functionals $E_2$ and $E_4$ over the SDiff orbits of BPS skyrmions was studied in Ref. \refcite{Sp2}. Due to the infinitely many dimensions of the space of SDiffs, this is, in general, a difficult problem, although it might be possible to find the minimizers in some simple cases with low baryon number. A first attempt to use this perturbative approach for the calculation of nuclear binding energies was undertaken in Refs. \refcite{Marl1,Marl2} based on the axially symmetric BPS skyrmion solutions (\ref{ax-sym}). Unfortunately, the axially symmetric configurations never minimize $E_2$ or $E_4$ (except for the spherically symmetric case $B=1$ - the hedgehog). In particular, $E_2$ grows like $B^\frac{7}{3}$ for large $B$ for the axially symmetric ansatz\cite{Sp2}, instead of a linear growth expected for true minimizers. As a result, the numerically fitted coefficients multiplying $E_2$ and $E_4$ result tiny in Refs. \refcite{Marl1,Marl2}, in order to be in agreement with the small binding energies of large nuclei.   

Having exposed the difficulties faced by a more accurate treatment of the binding energies problem, we shall restrict ourselves to a more modest goal, namely an approximate description of nuclear binding energies strictly within the BPS submodel, following Refs. \refcite{bind}. In this approach, the mass (static energy) of a nucleus $X$ is given by
\begin{equation}
E_X= E_{\rm sol} + E_{\rm sp} + E_{\rm is}+ E_{\rm C} + E_{\rm I}.
\end{equation}
Here, $E_{\rm sol}$ is the classical BPS soliton energy, $E_{\rm sp}$ and $E_{\rm is}$ are contributions from the semi-classical (rigid rotor) quantization of spin and isospin, $E_{\rm C}$ is the Coulomb energy contribution, and $ E_{\rm I}$ is due to a small explicit breaking of the isospin symmetry, which takes into account the mass difference between proton and neutron. Further, for simplicity, we will assume the axially symmetric ansatz (\ref{ax-sym}) for all the BPS skyrmion solutions we consider. This assumption enormously simplifies the calculations (e.g., the resulting baryon and energy densities are still spherically symmetric) and allows us to perform almost all calculations analytically. In particular, for axially symmetric skyrmions only two terms contribute in the expansion of the electric charge density into spherical harmonics, whereas there could be arbitrarily many terms for less symmetric skyrmions. The simplifying assumption of axial symmetry will introduce a certain error in our binding energy calculations, but this error will not affect all energy contributions in the same way. Concretely, both $E_{\rm sol}$ and $ E_{\rm I}$ are not affected at all by the symmetry assumption. Further, $E_{\rm sp}$ and $E_{\rm C}$ are probably not affected too much if the deviation of the corresponding physical nuclei (e.g., their baryon and energy densities) from spherical symmetry is not too pronounced. The biggest error will be produced for $E_{\rm is}$, so our results will not be able to reliably predict isospin excitational spectra. Therefore, we shall restrict our considerations to calculate, for each fixed baryon number $B$, the binding energy of the most abundant (supposedly also the most stable) nucleus. We will find that the resulting binding energies are quite close to their experimental values, especially for larger $B$, and we shall also explain why this is so. Before doing this, let us give a brief description of the calculation of $E_{\rm sp}$, $E_{\rm is}$, $E_{\rm C}$ and $ E_{\rm I}$.

\subsection{Semiclassical quantization of spin and isospin} 

Since spin and isospin are relevant quantum numbers of physical nuclei, this contribution constitutes an essential one. To proceed with the quantization, we introduce time-dependent rotations and iso-rotations around the static solitonic solution $U_0$, i.e.,

\begin{equation}
U(t,\vec x) = A (t) U_0 (R_B (t) \vec x) A^\dagger(t),
\end{equation}
with $R_B=\frac{1}{2} \Tr(\tau_i B \tau_j B^\dagger) \in {\rm SO(3)}$, and $A$ and $B$ being SU(2) matrices parametrized as $A(t) = a_0(t) + i a_i(t) \tau_i$ with $a_0^2 +\vec a^2 = 1$. After plugging this expression into the Lagrangian, we find

\begin{equation}
L_{\rm rot} = \frac{1}{2} \Omega_i \mathcal I_{ij} \Omega_j - \Omega_i \mathcal K_{ij} \omega_j + \frac{1}{2} \omega_i \mathcal J_{ij} \omega_j,
\end{equation}
with
\begin{equation} \label{iso-mom-int}
\mathcal I_{ij} = \frac{18 \lambda^2}{24^2} \int d^3 x \Tr(\epsilon^{pqr} T_i L_q L_r) \Tr(\epsilon^{pst} T_j L_s L_t),
\end{equation}
\begin{equation}
\mathcal K_{ij} = - \frac{18 \lambda^2}{24^2} \epsilon_{jkl} \int d^3 x \, x_k \Tr(\epsilon^{pqr} T_i L_q L_r) \Tr(\epsilon^{pst} L_l L_s L_t),
\end{equation}
\begin{equation}
\mathcal J_{ij} = \frac{18 \lambda^2}{24^2} \epsilon_{ikl} \epsilon_{jmn} \int d^3 x \, x_k x_m \Tr(\epsilon^{pqr} L_l L_q L_r) \Tr(\epsilon^{pst} L_n L_s L_t)
\end{equation}
being the moments of intertia tensors, $\omega_i$ and $\Omega_j$ the rotation and iso-rotation angular velocities, respectively, and $T_i = i U^\dagger [\frac{\tau_i}{2}, U]$.

The final step is to transform the generalized velocities into the canonical momenta and the Lagrangian into the Hamiltonian. Then, the nuclear quantum states will correspond to eingenstates of spin and isospin (Wigner $D$ functions):

\begin{equation}
|X \rangle = |j j_3 l_3 \rangle |i i_3 k_3 \rangle,
\end{equation}
where $X$ represents the nucleus, $\vec J$ ($\vec L$) is the space-fixed (body-fixed) angular momentum, $\vec I$ ($\vec K$) is the spaced-fixed (body-fixed) isospin angular momentum, and $j$, $j_3$, $l$, $l_3$ and $i$, $i_3$, $k$, $k_3$ are the corresponding eigenvalues.

Now we can use the axial symmetry of our ansatz to determine {\it a priori} the corresponding Hamiltonian (for the detailed calculation see Ref. \refcite{NayaThesis}; moments of inertia for axially symmetric configurations have been calculated in Ref. \refcite{hou-mag} for the standard Skyrme model). For this purpose, we have to distinguish the case with baryon number $B=1$ from $B > 1$ due to the fact that, for nucleons, the axial symmetry becomes spherical and, as is well known, there is an equivalence between spin and isospin for the hedgehog solution, and only one of the two constitutes an independent set of degrees of freedom (we choose spin, for concreteness). Therefore, the moments of inertia tensor is diagonal and proportional to the identity, $\mathcal J_{ij} = \mathcal J \delta_{ij}$, with
\begin{equation}
\mathcal J = \frac{4 \pi}{3} \lambda^2 \int dr \sin^4 \xi_r^2,
\end{equation}
whereas the Hamiltonian corresponds to the Hamiltonian of a spherical top (the body-fixed and space-fixed spin squared coindice),
\begin{equation}
\mathcal H_{\rm sp} = \frac{1}{2 \mathcal J} \vec L^2 = \frac{1}{2 \mathcal J} \vec J^2,
\end{equation}
with the static energy given in terms of the total spin quantum number $j$:
\begin{equation}
E_{\rm sp} = \frac{1}{2 \mathcal J} \hbar^2 j (j+1). 
\end{equation}
On the other hand, for nuclei with baryon number greater than one, we will have the Hamiltonian of a symmetric top due to the axial symmetry. It is characterized by the moments of inertia tensor $\mathcal J_{ij}= \mathcal J_i \delta_{ij}$, where $\mathcal J_1 = \mathcal J_2 = \mathcal J_3$ (the energy density is still spherically symmetric) but $\mathcal I_{ij}= \mathcal I_i \delta_{ij}$, where $\mathcal I_1 = \mathcal I_2 \not= \mathcal I_3$, concretely
\begin{equation}
\mathcal{I}_3 = |B|^{-\frac{1}{3}} \mathcal{J} \, ,\quad \mathcal{J}_1 = \mathcal{J}_2 = \mathcal{J}_3 = B^2 \mathcal{I}_3 \, , \quad \mathcal{I}_1 = \mathcal{I}_2 = \frac{3B^2 + 1}{4} \mathcal{I}_3 .
\end{equation}
The Hamiltonian reads
\begin{equation}
\mathcal H_{\rm sym-top} = \frac{L_1^2+L_2^2}{2 \mathcal J_1} + \frac{L_3^2}{2 \mathcal J_3} = \frac{\vec J^2}{2 \mathcal J_1} + \left( \frac{1}{2 \mathcal J_3} - \frac{1}{2 \mathcal J_1} \right) L_3^2.
\end{equation}
Thus, we expect to have two copies of it, one corresponding to the spin and another to the isospin. In addition, since a rotation of an angle $\phi$ about the three-axis can be undone by an iso-rotation of an angle $B\phi$ about the three-isospin axis, we will only take into consideration one of the corresponding generators ($L_3$ or $K_3$). Choosing $K_3$, the resulting static energy is
\begin{equation}
E_{\rm sp} + E_{\rm is} = \frac{\hbar^2}{2} \left( \frac{j(j+1)}{\mathcal J_1} + \frac{i(i+1)}{\mathcal I_1} + \left( \frac{1}{\mathcal I_3}-\frac{1}{\mathcal I_1} - \frac{B^2}{\mathcal J_1} \right) k_3^2 \right) .
\end{equation}
We remark that axial symmetry implies the relation $l_3 + Bk_3=0$ such that, to avoid unphysically large values of angular momentum, we should restrict to states with $k_3 =0$ which implies even $B$ (odd $B$ states have half-integer eigenvalues $k_3$), which we assume in the sequel.

Finally, it is also useful to have  explicit expressions for the canonical momenta as functions of the angular velocities, which from their definition read
\begin{equation}
\vec K = \frac{\partial L_{\rm rot}}{\partial \vec \Omega} = [\mathcal I_1 \Omega_1, \mathcal I_1 \Omega_2, \mathcal I_3 (\Omega_3 - B \omega_3)],
\end{equation}
\begin{equation}
\vec L = \frac{\partial L_{\rm rot}}{\partial \vec \omega} = [\mathcal J_1 \omega_1, \mathcal J_1 \omega_2, - B \mathcal I_3 (\Omega_3 - B \omega_3)].
\end{equation}

\subsection{Coulomb energy}

The Coulomb contribution is simply given by the usual generalization of the Coulomb energy to the volume charge density, that is to say,
\begin{equation}
E_{\rm C} = \frac{1}{2 \varepsilon_0} \int d^3 x d^3 x' \frac{\rho(\vec r) \rho(\vec r \, ')}{4 \pi |\vec r - \vec r \, '|},
\end{equation}
where $\rho$ is the expectation value of the electric charge densitiy with respect to the nucelar states $|X\rangle$. Indeed, the corresponding operator is given by \cite{Callan1984}
\begin{equation}
\hat \rho = \frac{1}{2} \mathcal B^0 + \mathbb J_3^0,
\end{equation}
with $\mathcal B^0$ being the baryon number density, and $\mathbb J_3^0$ the time-like component of the third isospin current density operator, $\mathbb J_3^\mu$, which reads \cite{Ding2007}
\begin{equation}
\mathbb J_3^0 = - \frac{i \lambda^2 \pi^2}{4} \epsilon^{0imn} \mathcal B_i \Tr \left[ \frac{\tau_3}{2} (\partial_m U U^\dagger \partial_n U U^\dagger - \partial_m U^\dagger U \partial_n U^\dagger U) \right],
\end{equation}
where $\mathcal B_i$ is the space-like component of the baryon current density.

Thus, after applying the semiclassical quantization to the electric charge density operator and using the axial symmetry of the ansatz, we arrive at \cite{bind}
\begin{equation}
\rho (\vec r)= \frac{1}{2} \mathcal B^0 + \langle X| \mathbb J_3^0 |X \rangle = - \frac{B}{4 \pi^2 r^2} \sin^2 \xi \xi_r + \frac{\lambda^2 i_3}{r^2 \mathcal I_3} \sin^4 \xi \xi_r^2 \frac{B^2+\cos^2 \theta}{3 B^2+1},
\end{equation}
where $i_3$ corresponds to the value of the third component of isospin. 

Once we have the expectation value of the electric charge density, we will use the multipole expansion of the Coulomb potential in order to calculate the corresponding contribution to the energy \cite{Carlson1963}:
\begin{equation}
\frac{1}{4 \pi |\vec r - \vec r \, '|} = \sum_{l=0}^\infty \sum_{m=-l}^l \frac{1}{2l+1} \frac{r^l_<}{r^{l+1}_>} Y^*_{lm} (\theta',\phi') Y_{lm} (\theta, \phi),
\end{equation}
where $r_< = \; {\rm min}(r,r')$ and $r_> = \; {\rm max} (r,r')$. Then, following the mechanism presented in Refs. \refcite{Marl2}, we expand the electric charge density into spherical harmonics, i.e.,
\begin{equation}
\rho (\vec r) = \sum_{l,m} \rho_{lm} (r) Y^*_{lm} (\theta,\phi),
\end{equation}
and defining the quantities
\begin{equation}
Q_{lm} (r) = \int_0^r dr' r'^{l+2} \rho_{lm} (r'),
\end{equation}
\begin{equation}
U_{lm} = \frac{1}{2 \varepsilon_0} \int_0^\infty dr r^{-2l-2} |Q_{lm}(r)|^2,
\end{equation}
the Coulomb energy is given by
\begin{equation}
E_{\rm C} = \sum_{l=0}^\infty \sum_{m=-l}^l U_{lm}.
\end{equation}
Within the considered ansatz, we only have two contributions to this expansion, namely
\begin{equation}
\rho (\vec r) = \rho_{00} (r) Y_{00} + \rho_{20} Y_{20,}
\end{equation}
with
\begin{equation}
\rho_{00}(r) = - \frac{B}{2 \pi^{3/2} r^2} \sin^2 \xi_r + \frac{2 \sqrt \pi}{3} \frac{\lambda^2 i_3}{r^2 \mathcal I_3} \sin^4 \xi_r^2,
\end{equation}
\begin{equation}
\rho_{20}(r) = \frac{4}{3} \sqrt{\frac{\pi}{5}} \frac{1}{3B^2+1} \frac{\lambda^2 i_3}{r^2 \mathcal I_3} \sin^4 \xi \xi_r^2.
\end{equation}
Thus, the Coulomb energy is 
\begin{equation}
E_{\rm C} = U_{00}+U_{20}.
\end{equation}
Again, we should have in mind that for $B=1$ the symmetry of the solution becomes spherical and only the $\rho_{00}$ remains with $E_{\rm C} = U_{00}$.

\subsection{Isospin breaking}

When plugging a specific solution into $\rho(\vec r)$, it can be seen that the Coulomb contribution produces a proton which is heavier than the neutron whilst nature tells us that it is the other way around. Therefore, the isospin symmetry is broken. To deal with this fact in a first approximation, we will consider the obvious leading order contribution, which is simply given by the Hamiltonian
\begin{equation}
\mathcal H_{\rm I} = a_{\rm I} I_3
\end{equation}
which commutes with the quantum operator $I_3$, such that $i_3$ remains a good quantum number, as must be true. Then, the corresponding energy reads
\begin{equation}
E_{\rm I} = a_{\rm I} i_3,
\end{equation}
where $a_{\rm I} < 0$ so a higher neutron mass can be accomplished.

\subsection{Results}

Now, in order to calculate the binding energies, first we must choose a specific potential and solve the BPS equation. Here we will use the square of the standard Skyrme potential 
\begin{equation}
\mathcal U = \mathcal U_\pi^2 =  \frac{1}{4}\Tr(1-U)^2,
\end{equation}
and the axially symmetric ansatz. In this case, and introducing the new coordinate
\begin{equation}
z=\frac{2\mu r^3}{3|B|\lambda}
\end{equation}
the BPS equation simplifies to
\begin{equation}
\sin^2 \xi \xi_z = - \sqrt{\mathcal{U}} = -1+\cos\xi
\end{equation}
(valid for all $B$) with the implicit solution
\begin{equation}
z= \pi - \xi - \sin \xi ,
\end{equation}
from which we directly get the solitonic energy through the BPS bound, whilst the additional contributions can be calculated as introduced above. However, since the total energy depends on the two parameters of our model plus the one from the isospin breaking (namely, $\mu$, $\lambda$ and $a_{\rm I}$), there is still one task left before comparing with experimental data: the determination of their numerical values. To proceed, we will fit our expression to three different quantities: the proton mass,
\begin{equation} \label{fit1}
M_p = 938.272 \; {\rm MeV};
\end{equation}
the experimental mass difference between neutron and proton,
\begin{equation} \label{fit2}
\Delta M = M_n - M_p = 1.29333 \; {\rm MeV};
\end{equation}
and the mass of a nucleus with magical numbers, concretely, $^{138}_{56}$Ba,
\begin{equation} \label{fit3}
M(^{138}_{56} {\rm Ba}) = 137.894 \; u,
\end{equation}
where $u = 931.494 \; {\rm MeV}$ is the unified atomic mass unit, so we also have to subtract the electron masses ($m_e= 0.511 \; {\rm MeV}$).

Thus, with the numerical value of the universal constants appearing in the calculations, i.e.,
\begin{eqnarray}
\hbar &=& 197.327 \; {\rm MeV} \; {\rm fm}, \\
\varepsilon_0 &=& \frac{1}{e} 8.8542 \cdot 10^{-21} \frac{1}{{\rm MeV} \; {\rm fm}}, \\
e &=& 1.60218 \cdot 10^{-19},
\end{eqnarray}
we arrive at the parameter values
\begin{equation}
\lambda \mu =  47.0563 \; {\rm MeV}, \quad \left( \frac{\mu}{\lambda} \right)^{1/3} = 0.536386 \; {\rm fm}^{-1}, \quad a_\textrm{I} = -1.65821 \; {\rm MeV}.
\end{equation}
The general definition of the binding energy of a nucleus $X$ is
\begin{equation}
\Delta E_{X} = Z E_\textrm{p} + N E_\textrm{n} - E_X,
\end{equation}
where $Z$ and $N$ are the number of protons and neutrons inside a nucleus $X$, $i_3 = (1/2)(Z-N)$, and $A\equiv B=Z+N$ is the usual nuclear physics notation for the atomic mass number. Our result for the binding energy then reads
\begin{eqnarray}
\Delta E_X (A,Z,j)&=& b_{{\rm V},A}A + b_{{\rm V},Z}Z - b_{\rm sp} j(j+1)A^{-\frac{5}{3}} \nonumber \\
&& - b_{\rm is,1}\frac{A^\frac{1}{3}}{1+3A^2} (A-2Z) - b_{\rm is,2}\frac{A^\frac{1}{3}}{1+3A^2} (A-2Z)^2 \nonumber \\
&& - b_{\rm C,1}A^\frac{5}{3} - b_{\rm C,2} A^\frac{2}{3} Z - b_{\rm C,3} A^{-\frac{1}{3}}Z^2 - b_{\rm C,4} \frac{(A-2Z)^2}{A^\frac{1}{3}(1+3A^2)^2}
\end{eqnarray}
where (all values are in MeV)
\begin{equation}
b_{{\rm V},A} = 9.881 \, , \quad b_{{\rm V},Z} = 0.3649 \, , \quad b_{\rm sp} = 13.174 \, , \quad b_{\rm is,1} = 26.35 \, , \quad b_{\rm is,2} = 13.174 \nonumber
\end{equation}
\begin{equation}
b_{\rm C,1} = 0.00072 \, , \quad b_{\rm C,2} = 0.00094 \, , \quad b_{\rm C,3} = 0.3639 \, , \quad b_{\rm C,4}= 0.00880
\end{equation}
In the binding energy expressions, both the classical soliton energies and the explicit isospin violating contributions cancel exactly. The two positive contributions $b_{V,A}$ and $b_{V,Z}$ stem from additional contributions to the nucleon energies. Concretely, $b_{V,A}$ receives the main contribution from the nucleon spin/isospin excitation, and a small contribution from the nucleon average Coulomb energy, whereas $b_{V,Z}$ gives the excess Coulomb energy of the proton. $b_{\rm sp}$ gives the spin contribution to the energy of the nucleus, the $b_{{\rm is},n}$ provide the isospin contributions, and the $b_{{\rm C},n}$ give the Coulomb contributions. 

\begin{figure}[t]
\begin{center}
\includegraphics[width=.8\textwidth]{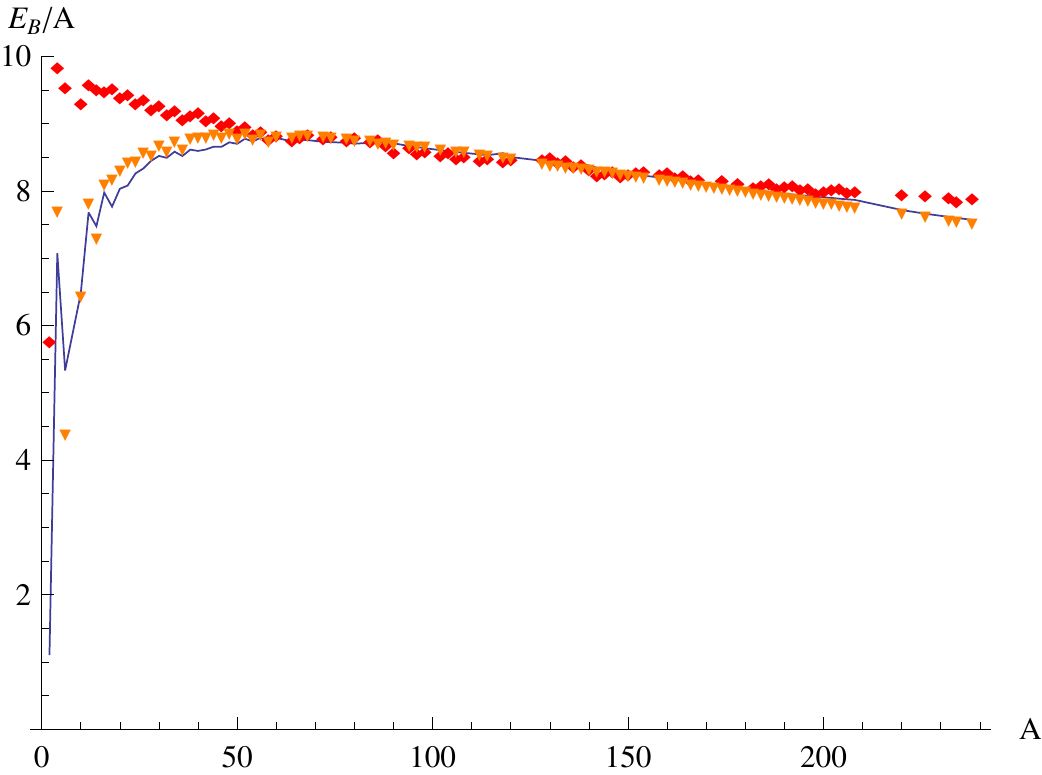}
\caption{Binding energies per nucleon in MeV. The experimental values are described by the solid line, our model results are represented by the diamonds, and the Weizs\"acker formula by triangles.}
\label{BE_Figure}
\end{center}
\end{figure}

It is useful to compare our results with the ones from the semi-empirical mass formula (Weizs\"acker formula) \cite{Krane}
\begin{equation}
 \Delta E_{X}^{\rm W}(A,Z) = a_{\rm V} A - a_{\rm S} A^{2/3} - a_{\rm C} Z (Z-1) A^{-1/3} 
 - a_{\rm A} \frac{(A-2Z)^2}{A} + \delta (A,Z), 
\end{equation}
where
\begin{eqnarray}
&& \nonumber \\ &&
\delta(n,Z) = \left\{ \begin{array}{cl}
a_{\rm P} A^{-3/4} & N \; \textrm{and} \; Z \; \textrm{even}, \\
0 & A \; \textrm{odd}, \\
- a_{\rm P} A^{-3/4} & N \; \textrm{and} \; Z \; \textrm{odd},
\end{array} \right.
\nonumber \\ && \nonumber \\ &&
a_{\rm V} = 15.5 \; {\rm MeV}, \quad a_{\rm S} = 16.8 \; {\rm MeV}, \quad a_{\rm C} = 0.72 \; {\rm MeV},  \nonumber
\\ &&
a_{\rm A}= 23 \; {\rm MeV}, \quad a_{\rm P} = 34 \; {\rm MeV} .  \nonumber 
\end{eqnarray}
We find that our term $b_{{\rm V},A}$ corresponds to the volume term $a_{\rm V}$ and our $b_{{\rm C},3}$ more or less corresponds to the Coulomb term $a_{\rm C}$ (there are no Coulomb self-energies for individual nucleons in the Weizs\"acker formula). Also, there are no spin contributions in the Weizs\"acker formula. As expected, nothing in the BPS submodel corresponds to the surface term $a_{\rm S}$ and to the pairing term $a_{\rm P}$. Our isospin term $b_{{\rm is},2}$ bears some similarity with the asymmetry term $a_{\rm A}$ in the Weizs\"acker formula, but the large $A$ behaviour is $\sim A^{-1}$ in the Weizs\"acker case but $\sim A^{-\frac{5}{3}}$ in our case. This is the announced too small isospin excitational energy related to the axial symmetry of our ansatz. Before further exploring this issue, we show our results for the binding energies per atomic weight number, together with the experimental values and the Weizs\"acker formula in Fig. \ref{BE_Figure}. 
Concretely, for each value of the atomic weight number $A$ we choose the values of $Z$ and $j$ corresponding to the most abundant nucleus.

We find that for small nuclei our model overestimates the binding energies. This is mainly because of the absence of a surface-like term and because in the BPS submodel, owing to its inherent collective character, single-nucleon properties are not described very well, both of which should improve with the inclusion of further terms (the near-BPS Skyrme model). Also the correct shapes of small $A$ nuclei (which, in general, will not be axially symmetric), should follow from this extended model. On the other hand, the BPS submodel describes the binding energies of the most stable large nuclei rather well, despite the problem with the too small isospin contributions. The reason for this is as follows. For the most stable nuclei, the Coulomb contribution and the asymmetry contribution ($\sim$ isospin contribution) to the binding energies are correlated (the valley of stability is defined by a balance condition between the two). A too small isospin contribution may, therefore, be compensated by a relatively bigger Coulomb contribution. This implies that either the Coulomb contribution will be slightly too big, or that the positive volume contribution $a_{\rm V}$ (from which both $a_{\rm C}$ and $a_{\rm A}$ are subtracted to reach the experimental binding energy) will be slightly smaller. We find that in our case, as a result of the fit (\ref{fit1}) - (\ref{fit3}), the second possibility is realised. Indeed, $a_{\rm V} \sim 15-16$ MeV in the Weizs\"acker formula, whereas $b_{{\rm V},A} \sim 10$ MeV in our model. 

We remark that the fact that the asymmetry term in the Weizs\"acker formula behaves like $A^{-1}(2Z-A)^2$ does {\em not} imply that the leading contribution to the isospin in the near-BPS model must behave exactly in the same way. After all, the (near-) BPS and Weizs\"acker binding energy formulae are different in many respects, and the BPS case even contains a further subleading isospin contribution. It is perfectly conceivable that for large $A$ the leading isospin contribution behaves like $A^{-\gamma}(2Z-A)^2$ for some $\gamma \in [1,\frac{5}{3}]$ where, however, the best value of $\gamma$ will probably be closer to 1 than to $\frac{5}{3}$. The value of $\gamma$ contains, in fact, an interesting physical information. From Eq. (\ref{iso-mom-int}) it follows that for a skyrmion configuration which is  a collection of (almost) non-overlapping, uncorrelated individual nucleons, the isospin moments of inertia behave like 
$\mathcal{I} \propto A$,
 which implies $\gamma =1$.  For our axially symmetric ansatz, on the other hand, the isospin moments of inertia are essentially equal to the spin moments of inertia, which results in $\gamma = \frac{5}{3}$, because 
$\mathcal{J} \propto A^\frac{5}{3}$ 
is the typical behaviour for spin moments of inertia. Physically, this corresponds to a state where individual nucleons are completely dissolved in a "nuclear soup" within the nucleus, and the values of isospin  in different regions of the nucleus are maximally correlated. The parameter $\gamma$, therefore, interpolates between maximally uncorrelated ($\gamma =1$) and maximally correlated ($\gamma = \frac{5}{3}$) nuclear matter, and its best fit value contains information about the amount of correlation of nuclear matter in nuclei.  

In any case, with our calculations of nuclear binding energies presented in this section we have probably gone as far as possible within an essentially analytical approach. We think that our results demonstrate the potential power of the near-BPS model as a reliabe EFT for nuclear matter, on the one hand, and the urgent necessity for a dedicated program of numerical investigation, on the other hand.

\subsubsection{Further potentials}

The quartic potential $\mathcal{U} = \mathcal{U}_\pi^2$ seems a simple and natural choice for the BPS potential, but obviously there are more possibilities. Besides, the potential $ \mathcal{U}_\pi^2$ is quite spiky close to the anti-vacuum $\xi = \pi$, so owing to the BPS equation, the baryon density is quite non-flat there, whereas the baryon density in nuclei is assumed not to vary too much, except close to the surface.
Therefore, we will now study the binding energies of nuclei with a new family of partially flat potentials defined as
\begin{equation} 
\mathcal U = \left \{ \begin{array}{ll}
U(\xi) & \xi \in [0,\xi_0], \\
1 & \xi \in [\xi_0,\pi],
\end{array} \right.
\end{equation}
\noindent where $U(\xi)$ is a non-flat contribution which can be considered as a skin or surface part of nuclei. Concerning this non-trivial part of the potential we will focus on a quartic approach to the vacuum. Thus, the specific expressions we will use are
\begin{equation} \label{part-flat}
\mathcal{U}_{{\rm pf}}(k,\xi) = \left \{ \begin{array}{ll}
\sin^4(k \xi) & \xi \in [0,\frac{\pi}{2 k}], \\
1 & \xi \in [\frac{\pi}{2k},\pi].
\end{array} \right.
\end{equation}
Note that by increasing $k$ we approach the limiting case of the step-function potential.

\begin{figure}[t]
\includegraphics[width=.45\textwidth]{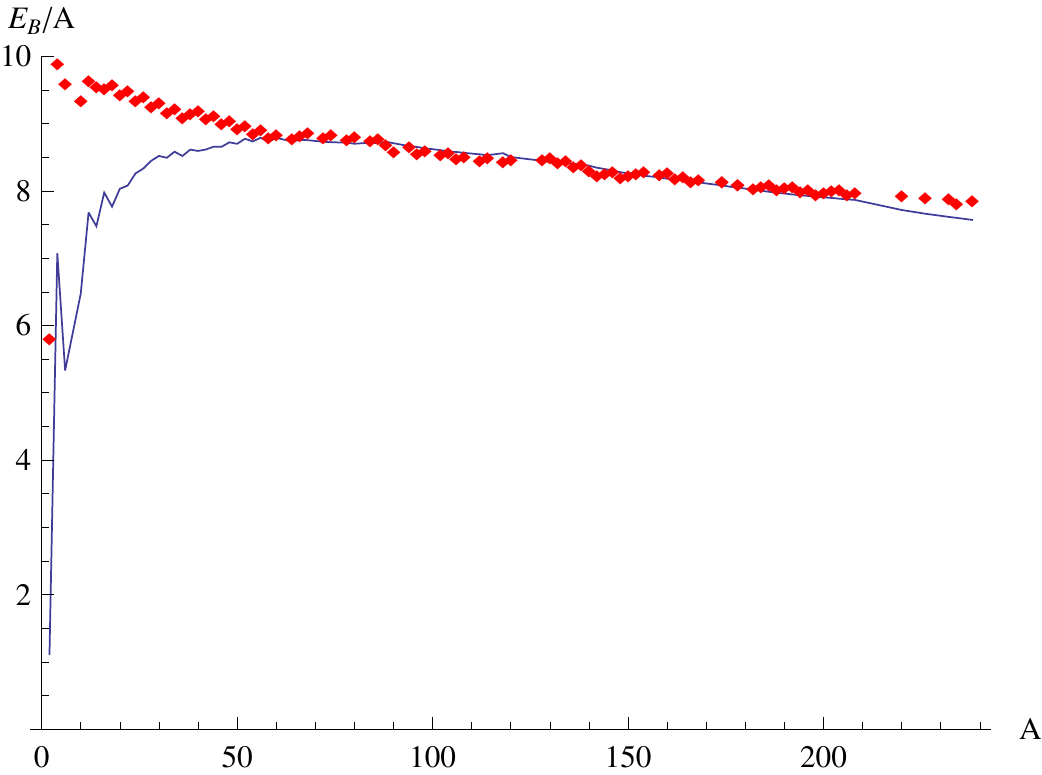}
\hspace{.5cm}
\includegraphics[width=.45\textwidth]{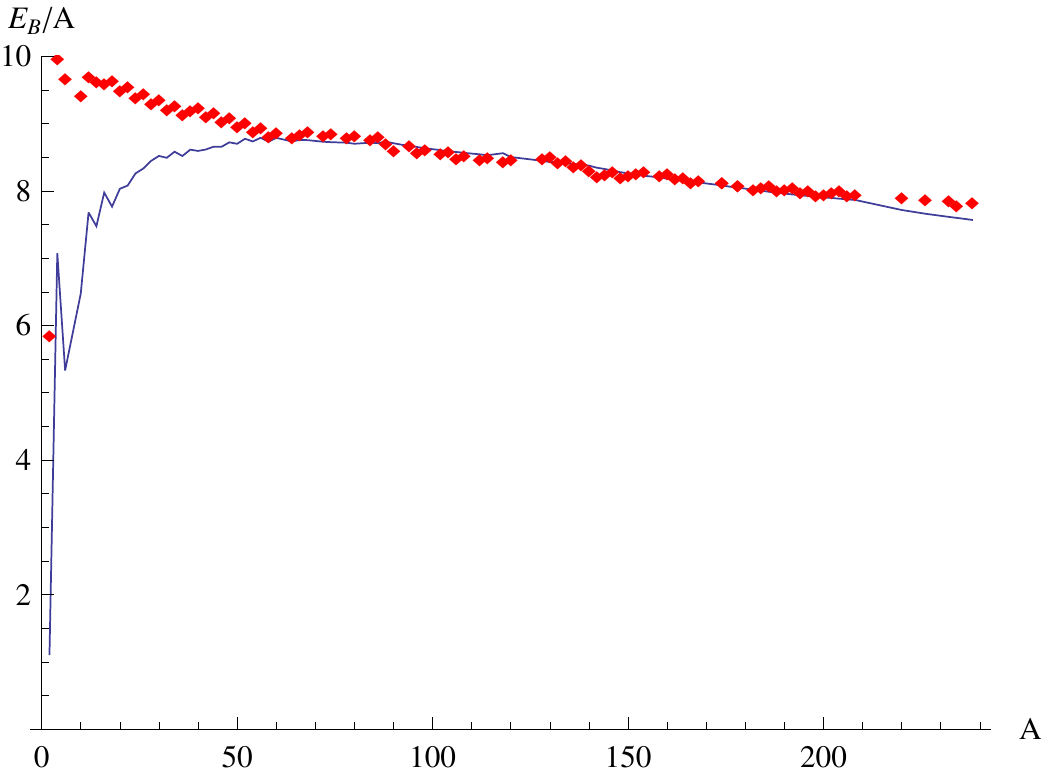} 
\caption{Binding energies per nucleon in MeV. Quartic partially flat potentials for $k=1$ (left) and $k=3/2$ (right).}
\label{StepLikePotentials}
\end{figure}

In order to calculate the binding energies we will proceed as previously indicated, with the same expressions for the additional contributions, but inserting the corresponding solution for each potential. Similarly, we will fit the parameters to the same masses as before. We present in Fig. \ref{StepLikePotentials} the results for two different partially flat potentials with $k=1, \frac{3}{2}$.

We find that the partially flat potentials lead to binding energy results which are very similar to the ones presented in the previous subsection for 
$ \mathcal{U}_\pi^2$. We remark that the results presented here are also very similar to the binding energy results calculated in Refs. \refcite{bind}, where
 the standard pion mass potential $ \mathcal{U}_\pi$ was used for simplicity, although - as explained here - this is not the proper choice. In other words, the binding energy calculations within the BPS Skyrme model only very weakly depend on the potential, the most important features being the BPS property together with the Coulomb, spin and isospin contributions.

\section{Neutron stars}\label{ns}
Up to now, we considered the near-BPS Skyrme model and its BPS limit as
 a model for nuclei and nuclear matter. After its coupling to gravity, the resulting self-gravitating Einstein-skyrmion system should then lead to a model of neutron stars (NS) for sufficiently large baryon number. In this section, we want to briefly describe self-gravitating BPS skyrmions and the resulting neutron stars and their properties, following Refs. \refcite{stars1,stars2}.  The coupling of skyrmions to Einstein gravity has been investigated, e.g., in Refs. \refcite{kli,bizon,piette1,piette2,walhout,piette3,nelmes,bjarke,gravSk}. In Ref. \refcite{piette3}, in particular, a MF EoS resulting from the skyrmion crystal in the standard Skyrme model in the large $B$ limit was coupled to gravity. The global properties of the resulting neutron stars are, in fact, qualitatively similar to the results we find for BPS Skyrme neutron stars (e.g., a similar $M(R)$ curve, although for a smaller maximum NS mass in the Skyrme crystal case), see Ref. \refcite{stars1} for a more detailed comparison.  

\subsection{Full field theory}

We shall continue to use the axially symmetric ansatz (\ref{ax-sym}), now in Schwarzschild coordinates defined by the spherically symmetric metric
\begin{equation} \label{metric}
ds^2 = {\bf A} (r) dt^2-{\bf B} (r) dr^2 - r^2 (d\theta^2 +\sin^2 \theta d\phi^2),
\end{equation}
leading to the spherically symmetric energy density and pressure expressions (see (\ref{perf-fluid}); $'\equiv \partial_r$)
\begin{equation} \label{rho-p-r}
\varepsilon = \frac{4B^2\lambda^2}{{\bf B} r^4} h(1-h) h'^2+\mu^2 \mathcal{U}(h), \;\;\; p=\varepsilon-2\mu^2 \mathcal{U}(h).
\end{equation}
Here, we introduced the new profile function
\begin{equation}
h=\frac{1}{2}(1-\cos \xi ) = \sin^2 \frac{\xi}{2}
\end{equation}
for convenience. The axially symmetric ansatz with the resulting spherically symmetric (energy, baryon and pressure) densities varying smoothly from a maximum value at the center ($r=0$) to zero at the neutron star radius $r=R$ are, in fact, much more realistic for neutron stars than for nuclei. Gravity tends to smooth deviations from spherical symmetry and to concentrate high energy density regions in the center, and the perfect fluid defined by the BPS Skyrme model does not resist such a rearrangement of its constituents. One might speculate that in the full near-BPS model some crystalline structure (or other types of local inhomogeneities) may survive close to the surface (in the neutron star crust) where the gravitational pull is weak, whereas a transition to a liquid phase essentially described by the BPS model occurs in the neutron star core (the strong-field region). This type of behaviour, in fact, reproduces the expected physical properties of neutron stars very well. Here we shall restrict calculations to the BPS submodel for simplicity, which still should describe the bulk (core) properties of neutron stars with reasonable accuracy.

Here, an important point is that the axially symmetric ansatz for the Skyrme field together with the Schwarzschild type ansatz for the metric are compatible with the Einstein equations 
\begin{equation} 
G_{\mu \nu} = \frac{\kappa^2}{2} T_{\mu \nu}
\end{equation}
(here $G_{\mu\nu}$ is the Einstein tensor and $\kappa^2 = 16 \pi G = 6.654 \cdot 10^{-41} \, {\rm fm} \, {\rm MeV}^{-1}$) 
and lead to a set of three ordinary differential equations: two for $h$ and ${\bf B}$ ($' \equiv \partial_r$), 
\begin{eqnarray}
\frac{1}{r} \frac{{\bf B}'}{{\bf B} } & =- & \frac{1}{r^2} ({\bf B} -1) +\frac{\kappa^2}{2} {\bf B}  \rho \label{eq1} \\
r({\bf B}  p)' &=& \frac{1}{2} (1-{\bf B} ) {\bf B}  (\rho +3p) +\frac{\kappa^2}{4} r^2 {\bf B}^2 (\rho-p)p \label{eq2} ,
\end{eqnarray}
and a third (decoupled) one determining ${\bf A}$ in terms of $h$ and ${\bf B}$, which we do not display here because it is not relevant for our purposes. The above system of two equations may be integrated by a shooting from the center, where the integration is done up to the radius $r=R$ where the pressure vanishes, $p(r=R)=0$, which defines the surface of the neutron star.  We refer to Refs. \refcite{stars1,stars2,NayaThesis} for a detailed discussion of the numerical integration procedure and the corresponding boundary conditions. Before a numerical integration can be done, we still have to choose numerical values for the two coupling constants $\lambda$, $\mu$ of our model. As here we only consider classical self-gravitating soliton solutions (no quantum or isospin or Coulomb corrections), we fit to infinite nuclear matter. That is to say, for each potential we choose the fit values for $\lambda$ and $\mu$ such that the (non-gravitating) BPS skyrmions reproduce the binding energy and saturation density  
(\ref{nuc-sat-values}) of infinite nuclear matter. 

\subsection{The TOV approach}
The approach presented so far amounts to a complete field-theoretic calculation of self-gravitating (BPS) skyrmions, where the gravitational backreaction on the matter (Skyrme) field has been fully taken into account. This is, however, {\em not} the way neutron stars are calculated from standard nuclear effective field theories (EFTs). Indeed, any attempt to directly solve the Einstein-EFT system for a huge ($\sim 10^{57}$) number of nucleons is hopeless. Instead, a kind of averaging procedure leading to a macroscopic (thermodynamical) description is required. Typically, in EFTs this averaging is provided by a mean-field (MF) approximation, where elementary or composed field operators are replaced by their average (in-medium) expectation values. The resulting densities (without gravity) are, therefore, necessarily constant, and, provided that the macroscopic description leads to a simple perfect fluid, this fluid is barotropic by construction, i.e., the (spatially constant) average energy density $\bar \varepsilon$ and the (spatially constant) average pressure density $\bar p$ are related by an algebraic relation (equation of state) 
$\bar \varepsilon = \bar \varepsilon (\bar p)$. If this perfect fluid is coupled to gravity and a static, spherically symmetric metric (\ref{metric}) is assumed, then consistency requires that the densities, too, depend on the Schwarzschild coordinate $r$ (they cannot be constant, because the general-relativistic hydrostatic equilibrium must balance the gravitational pull). The resulting system of Einstein equations is then formally completely equivalent to the system (\ref{eq1}), (\ref{eq2}). The only difference is that $\varepsilon$ and $p$, which are certain functions of the field variables (here, the Skyrme field) and their first derivatives, i.e., derived quantities in terms of the true, microscopic degrees of freedom, are replaced by the macroscopic, effective MF degrees of freedom $\bar \varepsilon$ and $\bar p$. Besides, the macroscopic MF variables $\bar \varepsilon$ and $\bar p$ are considered independent in Eqs. (\ref{eq1}), (\ref{eq2}), such that the MF EoS  $\bar \varepsilon = \bar \varepsilon (\bar p)$ is needed as a third equation to close the system. In contrast, for the true, microscopic densities, the system of Einstein equations (\ref{eq1}), (\ref{eq2}) closes by itself, and no further information is needed. The two equations (\ref{eq1}), (\ref{eq2}) for two independent variables $\bar \varepsilon$ and $\bar p$ toghether with the EoS $\bar \varepsilon (\bar p)$ are called the Tolman-Oppenheimer-Volkoff (TOV) \cite{OV,Tol} equations. 

As discussed at length in Section 3, a MF limit may be performed and the corresponding average densities may be easily calculated in the BPS Skyrme model. In addition to a full field-theoretic calculation of gravitating skyrmions we may, therefore, calculate them within the TOV approach, using the MF EoS (\ref{MFEoS}), where the constant $P$ must be replaced by the MF pressure density $\bar p$. On the one hand, such a MF TOV calculation is completely analogous to neutron star calculations in other EFTs and, therefore, facilitates the comparison with these approaches. On the other hand, the MF TOV calculation gives us the rare opportunity to compare exact field theory and MF results and to estimate the possible error introduced by the latter. As full field theory calculations of self-gravitating nuclear matter are impossible in other approaches, this possibility is a unique feature of the BPS Skyrme model.

\subsection{Results} 
Before presenting the results of our numerical integration, we want to briefly explain a small technical difference between the exact and the MF TOV case. In the MF TOV case, for each BPS Skyrme model (each choice of a potential, which leads to a given MF EoS $\bar \varepsilon (\bar p)$), we may choose $\bar \varepsilon_0 \equiv \bar \varepsilon (r=0)$ as an initial value for the shooting procedure such that each initial value leads to a neutron star solution with a certain radius $R(\bar\varepsilon_0)$ and mass $M(\bar\varepsilon_0)$ (the neutron star is stable as long as $(\partial M/\partial \bar \varepsilon_0) \ge 0$). As a consequence, all bulk quantities characterizing a neutron star [its radius $R$, mass $M$ and baryon number (non-gravitational mass) $B$] are derived quantities (they must be derived from a particular solution). Explicitly, $R$ is defined by $\bar p(R)=0$, whereas $M$ and $B$ are defined by
\begin{equation} \label{NS-mass}
M=4\pi \int_0^R dr r^2 \bar \varepsilon (r) ,
\end{equation}
\begin{equation}
B = 4\pi \int_0^R dr r^2 \sqrt{\bf B} \bar \rho_{B} (r) 
\end{equation}
where $\bar \rho_B$ is defined in Eq. (\ref{bar-rho}).
In the exact field theory calculation, on the other hand, the energy density $\varepsilon (r=0)\equiv \varepsilon_0$ at the center (equivalently, the second Taylor coefficient of $h$ in an expansion about $r=0$), is not a free parameter, because regular solutions must obey an additional condition at the neutron star surface (concretely, $p'(R)=0$ in addition to $p(R)=0$). To find a physically acceptable solution, $\varepsilon_0$ must therefore be varied till this condition is met. In this case, different neutron star solutions may be found because baryon number (non-gravitational mass) is no longer a derived quantity but, instead, a free parameter in the axially symmetric ansatz, see Eq. (\ref{rho-p-r}), such that different values of $B$ lead to different solutions. Here, solutions cease to exist beyond a certain maximum value $B_{\rm max}$, and the solution for $B_{\rm max}$ defines the maximum neutron star mass $M_{\rm max}$ for each model (each choice of the potential). For more details we refer to \cite{stars1,stars2}.    

Concretely, we shall consider the two potentials $\mathcal{U} = \mathcal{U}_\pi^2 = 4h^2$ and the partially flat potential (\ref{part-flat}) for $k=1$, leading to the fit values
\begin{equation}
\mathcal{U}_\pi^2 \; : \quad \lambda^2 = 15.49\; {\rm MeV} \; {\rm fm}^3 \, , \quad
\mu^2 = 141.22 \; {\rm MeV} \; {\rm fm}^{-3}
\end{equation}
\begin{equation} \label{part-flat-k=1}
\mathcal{U}_{\rm pf} (k=1)  \; : \quad \lambda^2 = 23.60 \; {\rm MeV} \; {\rm fm}^3 \, , \quad
\mu^2 = 121.08 \; {\rm MeV} \; {\rm fm}^{-3}
\end{equation}

\begin{figure}[t]
\begin{center}
\includegraphics[width=.8\textwidth]{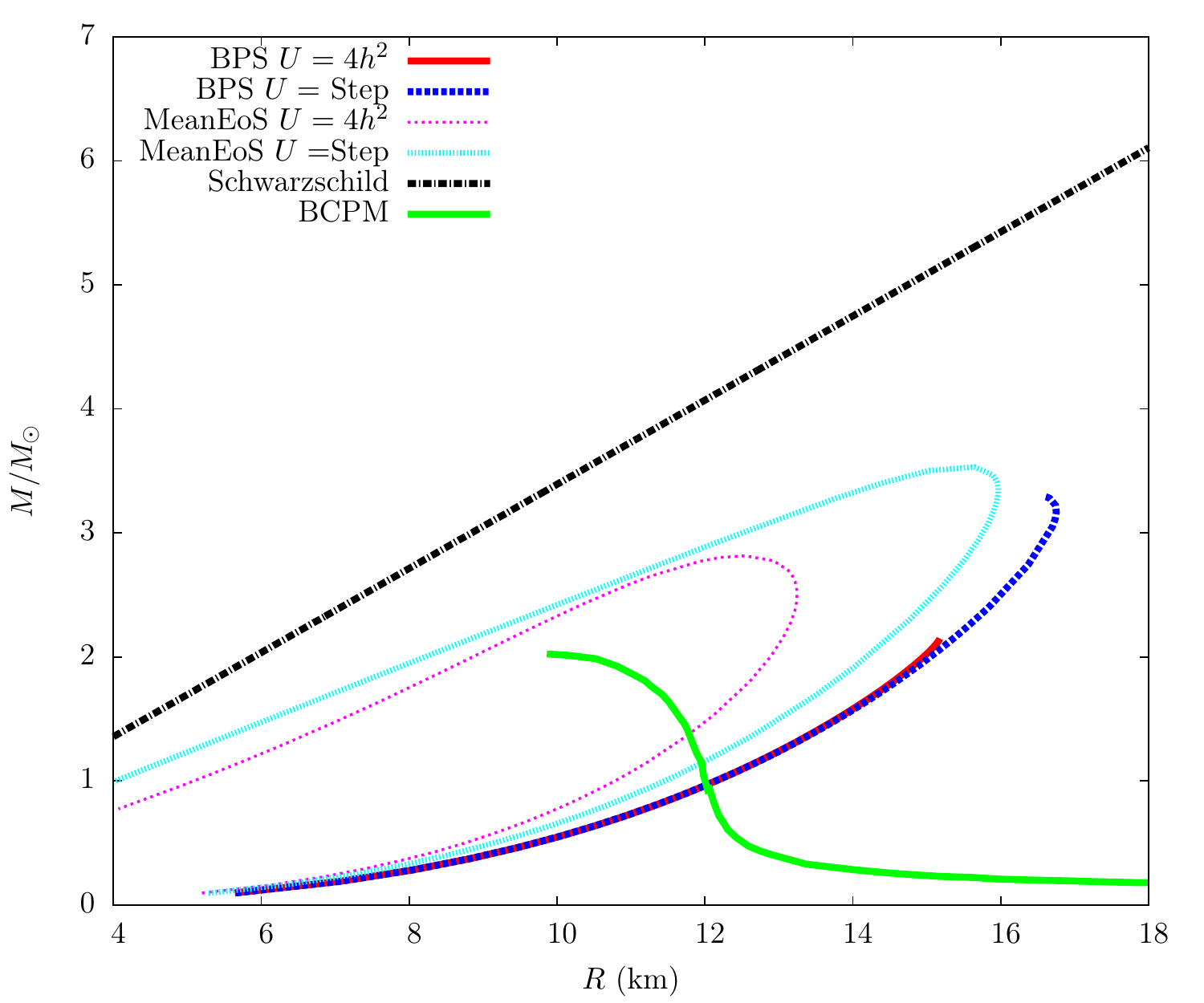}
\caption{Neutron star masses in solar units as a function of the neutron star radii (in km), both for the potential $\mathcal{U}_\pi^2= 4h^2$ and for the partially flat potential (\ref{part-flat-k=1}) (here called "Step"). We show both the exact case "BPS" and the MF TOV calculation "MeanEoS". For comparison, we also show the $M(R)$ curve "BCPM" resulting from the EoS of Ref. \refcite{vinas}.}
\label{mass}
\end{center}
\end{figure}

In Fig. \ref{mass} we show the masses and radii of the resulting neutron star solutions (i.e., the $M(R)$ curves) for each model, both for the exact and for the MF TOV calculations. In the MF case, the "initial value" $\bar \varepsilon_0$ grows along the curve, so the stable branch is the branch of growing $M$ (the lower branch up to $M_{\rm max}$; we also show the unstable branch for simplicity).  For both potentials, a certain difference between exact and MF calculations may be appreciated, the MF calculations always leading to more compact neutron stars (smaller radius for the same mass) and to bigger maximum masses. We remark that for the standard pion mass potential $\mathcal{U}_\pi$ (which we do not consider here for the reasons explained earlier), the difference between MF and exact calculation is much smaller, see \cite{stars2}. This seems to imply that for this difference the approach to the vacuum (quartic vs. quadratic in our particular examples) is more important than the (more or less flat) behaviour at the center (close to the anti-vacuum). 
We further find that the masses grow with the radius, except very close to the maximum mass.
For the sake of comparison, we also plot the TOV calculation result for a representative nuclear physics EoS derived using standard nuclear physics methods, concretely the Brueckner-Hartree-Fock method, together with a density functional method (the so-called BCPM nuclear energy functional) to describe inhomogeneities relevant for the neutron star crust \cite{vinas}. The resulting $M(R)$ curve is rather different from our curves, and we shall further comment on these differences in the discussion subsection.

\begin{figure}[t]
\begin{center}
\includegraphics[width=.8\textwidth]{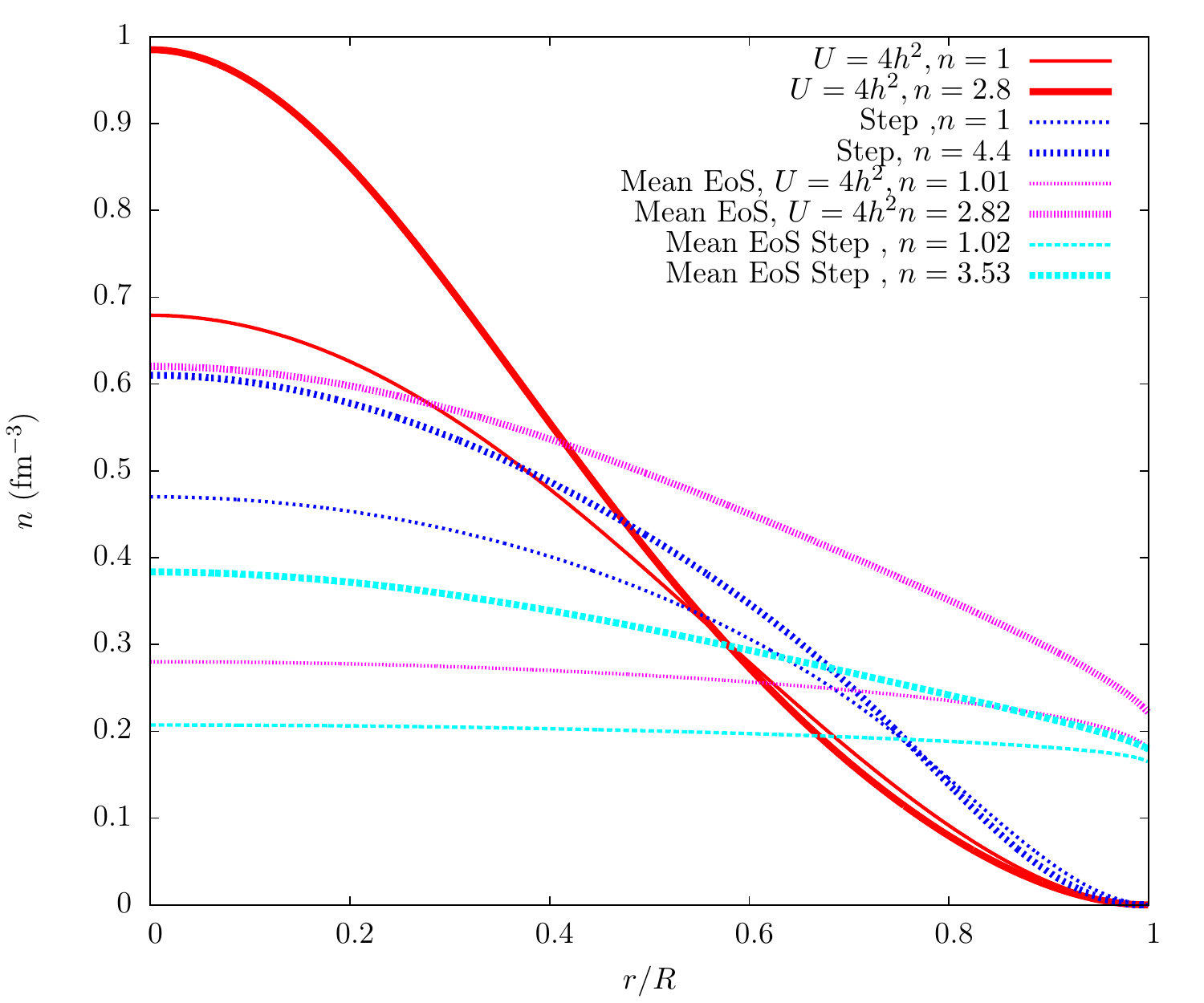}
\caption{Baryon number density (in fm$^{-3}$), as a function of the radius (in units of the corresponding neutron star radius), for the potential $\mathcal{U}_\pi^2= 4h^2$ and for the partially flat potential (\ref{part-flat-k=1}) (here called "Step"). We show both the exact case and the MF TOV calculation "MeanEoS", for different values of the baryon number in solar units, $n=B/B_\odot$.}
\label{baryon-vs-r}
\end{center}
\end{figure}

In Fig. \ref{baryon-vs-r} we plot the baryon number densities. The MF densities show less variation over the whole NS radius and take the nonzero value $\bar \rho_B(R)=\rho_s$ at the surface. In the plot, they do not reach exactly the surface value because, for numerical reasons, we cut the plot slightly before $\bar p=0$. More concretely, the reason is that the $\bar \rho_B (r)$ curves get much steeper close to the surface, because the MF EoS is much softer there (zero MF speed of sound at $\bar p=0$, see Section 3.4).
The exact densities vary from a much bigger central value $\rho_B(0)$ to $\rho_B (R)=0$ at the surface, again manifesting the non-barotropic nature of the corresponding fluid. This behaviour does {\em not} mean that the exact densities are compressed a lot by gravity in the center. The exact baryon densities have similar, bell-shaped graphs already in the case without gravity (for BPS skyrmions), and the additional compression induced by gravity never exceeds a factor of about two, even for the most massive NS. This is related to the stiff character of (BPS) skyrmionic matter.

\begin{figure}[t]
\begin{center}
\includegraphics[width=.8\textwidth]{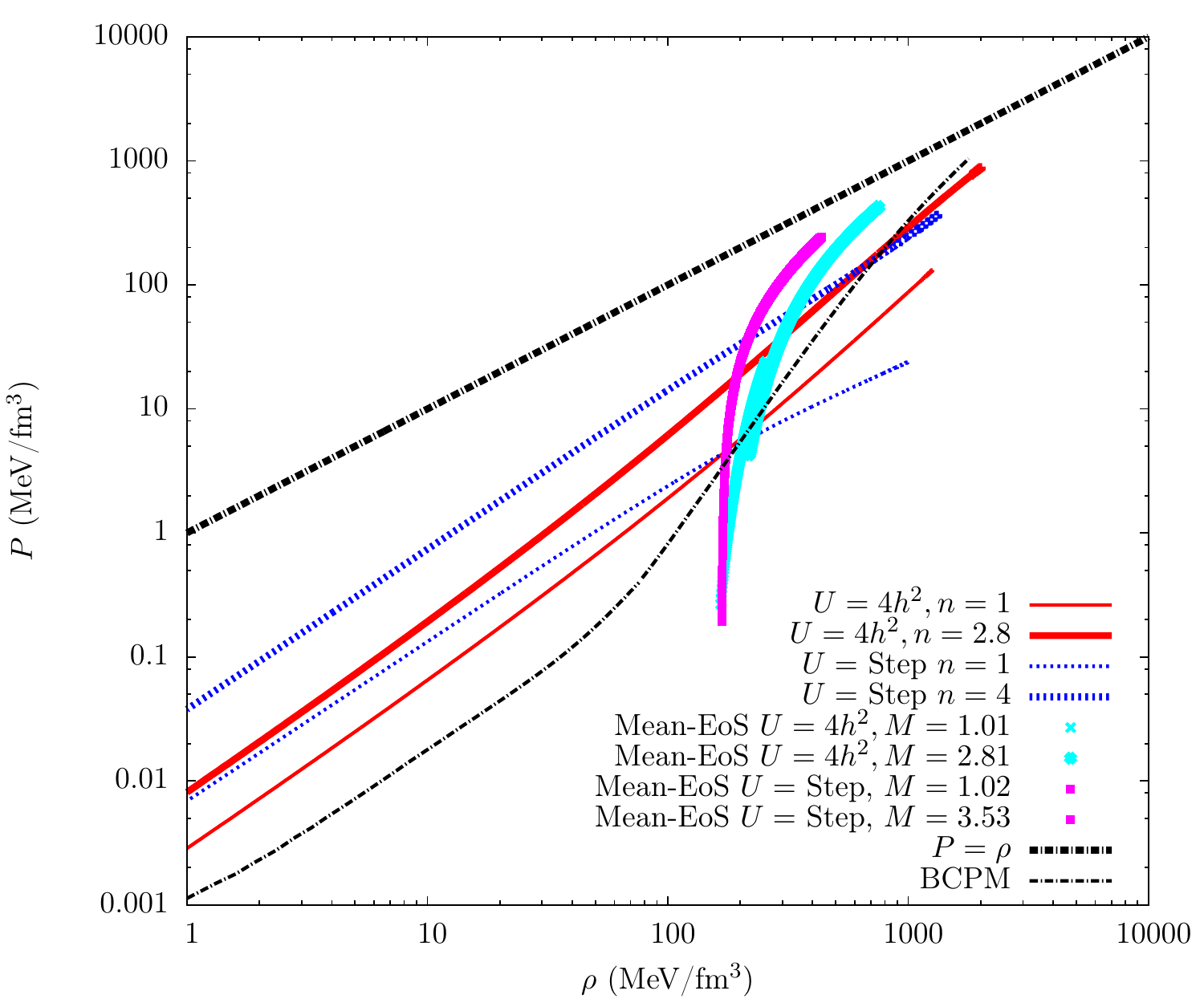}
\caption{On-shell EoS $\bar p(\bar \varepsilon)$, for the potential $\mathcal{U}_\pi^2= 4h^2$ and for the partially flat potential (\ref{part-flat-k=1}) (here called "Step"). We show both the exact case and the MF TOV calculation "MeanEoS", for different values of the baryon number in solar units, $n=B/B_\odot$.We also show the EoS "BCPM" from \cite{vinas} for comparison.}
\label{onshell-eos}
\end{center}
\end{figure}

In Fig. \ref{onshell-eos} we show the on-shell EoS for several NS solutions. By "on-shell EoS" we mean the following. Each NS solution leads to a function $\varepsilon (r)$ ($\bar \varepsilon (r)$) and to a function $p(r)$ ($\bar p(r)$). By eliminating the independent variable $r$ from this pair of functions, we may then construct curves $p(\varepsilon)$ ($\bar p (\bar \varepsilon)$) at least numerically, which are our on-shell EoS. In the case of the MF TOV calculation, all solutions for the same model (the same potential) have the same MF EoS, therefore all their on-shell EoS must coincide with each other and with this MF EoS. And indeed, we find precisely this behaviour in Fig. \ref{onshell-eos}. For the full field theory calculation, on the other hand, a barotropic EoS does not exist. Different solutions will, therefore, lead to different on-shell EoS, even for the same potential. Numerically it turns out that these on-shell EoS for the full field theory are rather well described by the EoS of a polytrope, $p \sim a \varepsilon^b$ where, however, the numbers $a$ and $b$ are different for different solutions, even for the same potential.

\begin{figure}[t]
\begin{center}
\includegraphics[width=.8\textwidth]{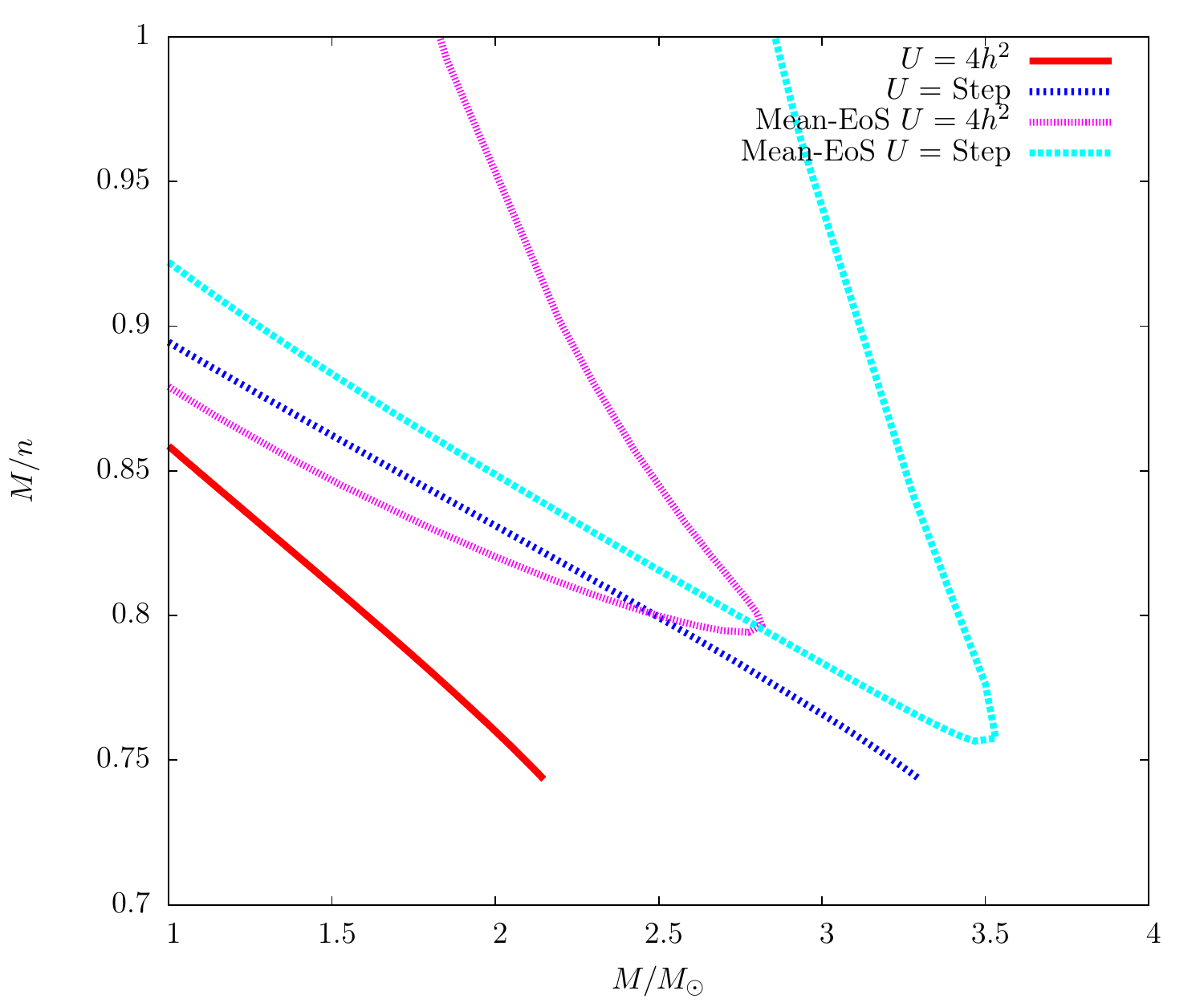}
\caption{Gravitational mass loss. Vertical axis: $((M/M_\odot)/(B/B_\odot))$. Horizontal axis: $(M/M_\odot)$. For simplicity, we also show the unstable branches in the MF TOV case.}
\label{massloss}
\end{center}
\end{figure}

Finally, in Fig. \ref{massloss} we plot the gravitational mass loss of our different NS solutions. For the exact calculations, the maximum mass loss is about 25\% for both potentials.

\subsection{Discussion}
Our results have several important implications for neutron stars, both within and beyond Skyrme models. Here we want to highlight the most important ones.
\begin{itemize}
\item
One first observation which meets the eye concerns the rather different $M(R)$ curves between many standard nuclear physics results and the BPS Skyrme model, see Fig. \ref{mass}. The basic reason for this difference is the rather stiff EoS of the BPS Skyrme model (in the following discussion we shall restrict to the MF version of the BPS model, because the existence of a barotropic EoS facilitates comparisons). Indeed, it can be seen easily in Fig. \ref{onshell-eos}, that, e.g., the BCPM EoS \cite{vinas} is much softer than the MF EoS of the BPS model except in the limit of very high densities. The BPS model MF EoS, in fact, rapidly approaches the maximally stiff EoS $\bar \varepsilon = \bar p \; + $ const. (which is called the "maximally compact" EoS in the neutron star literature), and this maximally compact EoS is known to lead to $M(R)$ curves very similar to ours. 

To judge the physical meaning of this difference, one should distinguish the case of light neutron stars from the heavier ones. For light neutron stars, the low-density part of the EoS corresponding to the NS crust will provide an appreciable contribution to the total NS radius. Further, these low-density EoS are rather well-known from standard nuclear physics. The BPS Skyrme model, on the other hand, does not describe surface (crust) contributions, so any low-density completion of the BPS model should lead to a tail of low-density nuclear matter and, therefore, to larger NS radii. It should be remarked, however, that low mass neutron stars are not firmly established observationally, so the discussion about the "true" $M(R)$ curves for low-mass neutron stars could well be a rather academic one. For higher mass neutron stars, pressure will rapidly grow towards the inside of the star to values where the low-density EoS no longer applies, and the low-density contribution to masses and radii is negligible. In the medium density regime (for $1.2 \rho_s \le \bar \rho_B \le 3\rho_s$, say), the EoS of the BPS model is still much stiffer than most nuclear matter EoS. In this regime, however, the BPS model should already lead to a reasonable description of nuclear matter within the Skyrme model approach. Further, also the skyrmion crystal of the standard Skyrme model \cite{piette3} leads to rather similar $M(R)$ curves. And indeed, although the EoS of the skyrmion crystal is softer than the EoS of the BPS model, in the medium density regime it is still stiffer than typical nuclear matter EoS from more traditional approaches. The skyrmion crystal $M(R)$ curves are, in fact, very similar to the $M(R)$ curves of compact quark stars, because their asymptotic large density EoS coincide. Both the standard Skyrme model EoS \cite{eos} and the quark matter EoS approach   
the ultrarelativistic free fermion EoS $\bar \varepsilon = 3 \bar p + \ldots$ in this limit. 

In other words, for sufficiently massive NS (above 1.5 solar masses, say) the behaviour $(dM/dR)>0$ of the $M(R)$ curve for most NS masses (except, probably, very close to $M_{\rm max}$) is a genuine prediction of Skyrme models which is related to the rather high stiffness of Skyrme matter. This behaviour is different from the predictions of many traditional nuclear physics models, \cite{latt,ozel,piek}  but is perfectly compatible with the (still rather scarce) observational data.  

\item
A second, related observation is that the maximum NS masses of the BPS Skyrme model are easily compatible with some recent observations of  rather massive neutron stars \cite{latt} of up to $2.5 M_\odot$, which many traditional nuclear physics models have difficulties to accomodate.

\item
Another important issue concerns the difference between full field theory and MF calculations within the BPS Skyrme model. Already for bulk quantities like masses or radii there exist appreciable differences, see Fig. \ref{mass}. In general, the MF neutron stars tend to be more compact and to allow for slightly higher maximum masses than their full field theory counterparts. If this difference between full field theory and MF $M(R)$ curves happens to occur also in other models, this causes a serious problem for the so-called TOV inversion procedure. Indeed, for a given EoS $\bar \varepsilon (\bar p)$ the TOV equations determine a unique $M(R)$ curve. It is, therefore, possible to invert this procedure and to derive an EoS $\bar \varepsilon (\bar p)$ from a given $M(R)$ curve.\cite{lind,stein} 
But the whole construction hinges, of course, on the assumption that a barotropic EoS - probably resulting from a MF approximation - provides a sufficiently faithful description of the underlying EFT of nuclear matter. If the MF approximation introduces a certain error such that the $M(R)$ curve of the full EFT differs from the curve for the MF approximation, then the TOV inversion will reconstruct a ficticious EoS which is unrelated to the MF EoS of the original EFT. 

For local densities, the difference between exact and MF calculations is much more pronounced. They are, in fact, completely different, see, e.g., Fig. \ref{baryon-vs-r}. This implies that certain observables which depend on the local distribution of energy or particle number like, e.g., the moments of inertia relevant for the description of rotating neutron stars, will probably differ quite a lot between full field theory and MF results.

\item 
In addition to TOV calculations for specific EoS, there exist some (rather) EoS-independent bounds in the neutron star literature. In our BPS model, we still have the freedom to choose different potentials, and for a more accurate description the model should be extended, e.g., to the full near-BPS model. Therefore, precise quantitative predictions are still premature. It may, nevertheless, be of some interest to compare the results of the BPS model with the generic bounds just mentioned (for details, we refer to Ref. \refcite{stars2}). The first bound is the so-called Rhoades-Ruffini bound\cite{RR,glend} on the maximum neutron star mass. The bound is derived by joining a well-established nuclear physics EoS for $0\le \bar \rho_B \le \rho_f$ for some $\rho_f > \rho_s$ with the maximally stiff EoS for $\bar\rho_B >\rho_f$. By assuming a plausible value for $\rho_f$, Rhoades and Ruffini found $M_{\rm max} \le 3.2 M_\odot$\cite{RR}. By decreasing the value of $\rho_f$, the bound may be weakened. In particular, for the limiting case $\rho_f = \rho_s$ one finds the weaker bound $M_{\rm max} \le 4.3M_\odot$\cite{glend}. In a similar fashion, a bound for the compactness of a NS may be derived. The compactness parameter is defined as
\begin{equation}
\beta = \frac{2GM}{R} \equiv \frac{R_S}{R}
\end{equation}
where $R_S \equiv 2GM$ is the Schwarzschild radius. Obviously, $\beta \le 1$. Using the same assumptions that Rhoades and Ruffini used for their mass bound, Glendenning derived the bound\cite{glend1}
\begin{equation} \label{glend-bound}
\beta \le \frac{1}{1.47} = 0.68 \equiv \beta_{\rm G}.
\end{equation}
Finally, Lattimer and Prakash\cite{latt1} proposed a phenomenological bound on the central energy density $\varepsilon (r=0)\equiv \varepsilon_c$ in terms of the central energy density of the (very compact) Tolman VII solutions for a compactness equal to the Glendenning bound, concretely
\begin{equation} \label{rho-c-bound}
 \varepsilon_c \le \left( \frac{\beta_{\rm G}}{2G}\right)^3 \frac{15}{8\pi  } \frac{1}{M^2} \equiv \varepsilon_{c,{\rm TVII}} \simeq 1.45 \cdot 10^{19} \left( \frac{M_\odot}{M} \right)^2 {\rm kg}\, 
 {\rm m}^{-3} .
 \end{equation}
It turns out that almost all nuclear physics EoS used for the description of neutron stars obey this phenomenological bound.

In Table 1, we show the corresponding values of our model for the two potentials considered, both for the exact and the MF calculations. We find that for the partially flat potential the maximum mass is slightly above the Rhoades-Ruffini bound even for the exact calculation. This just means that the corresponding skyrmionic matter gets quite stiff rather soon, and for a generalized Rhoades-Ruffini bound implies a value of $\rho_f$ which is rather close to the nuclear saturation density $\rho_s$. The Glendenning bound on $\beta$ is satisfied by both potentials. Finally, the central energy density for the exact maximum mass solution for the potential $4h^2$ is slightly above the corresponding Tolman VII value. 

\begin{table}[ht]
\tbl{Several observables for maximum mass neutron stars, both for the partially flat (=p.f.) potential and for the potential $\mathcal{U}_\pi^2$.  }
{\begin{tabular}{@{}ccccc@{}} \toprule
 Potential  & p.f., exact & p.f., MF & $4h^2$, exact & $4h^2$, MF \\
\colrule
$M_{\rm max}/M_\odot$ & 3.29  & 3.53 & 2.15 &2.82 \\
$B_{\rm max}/B_\odot$ &4.43 & -&2.89 & - \\
$\beta$ for $M_{\rm max}$  & 0.59  & 0.67 & 0.42 &0.66 \\
$\varepsilon_c/(10^{18} \, {\rm kg}\, {\rm m}^{-3})$ for $M_{\rm max}$  & 1.20  & 0.775 & 4.01 &1.34 \\
$ \varepsilon_{c,{\rm TVII}}/(10^{18} \, {\rm kg}\, {\rm m}^{-3})$  & 1.34  & 1.16 & 3.14 &1.82 \\
\botrule
\end{tabular}
}
\begin{tabnote}
\end{tabnote}
\label{table1}
\end{table}

\item
Finally, we want to compare our full field theory $M(R)$ curves with some particular constraints discussed in Ref. \refcite{blaschke}. In Fig. \ref{constraints}, we show these constraints together with our $M(R)$ curves and the one from the EoS "DBHF" of Ref. \refcite{blaschke} which meets all the constraints. The following constraints are shown. i) A mass estimate for the neutron star 4U 1636-536 of $M=(2.0\pm 0.1)M_\odot$\footnote{We remark that the currently accepted best high-mass constraints come from the neutron stars PSR J1614-2230 ($M = (1.97\pm0.04)M_\odot$)\cite{demorest} and PSR J0348+0432 ($M=(2.01\pm0.04)M_\odot$)\cite{antoniadis}. These are, however, very close to the older mass constraint of 4U 1636-536, therefore we decided to reproduce the original constraints in Fig. \ref{constraints} exactly as shown in Ref. \refcite{blaschke}.}. ii) The high-frequency brightness oscillation measurements of the neutron star 4U 0614 +09 which constrain the mass $M$ and radius $R$ of the NS to a certain wedge-shaped region. iii) A constraint in the $M$-$R$ plane stemming from the measurement of the thermal radiation of the neutron star RX J1856, together with an estimate of its distance. For details we refer to Ref. \refcite{blaschke}. We find that both our curves and the curve for the DBHF model satisfy all bounds. We remark that many models have difficulties in satisfying the first bound (a NS with a mass of $\sim 2M_\odot$) and, especially, the third bound (from the neutron star RX J1856) which requires rather large radii and/or large masses, so the fact that our model easily accomodates both of them is an interesting observation. Of course, for small masses our curves will change after a more complete treatment, as already explained.

\begin{figure}[t]
\begin{center}
\includegraphics[width=.8\textwidth]{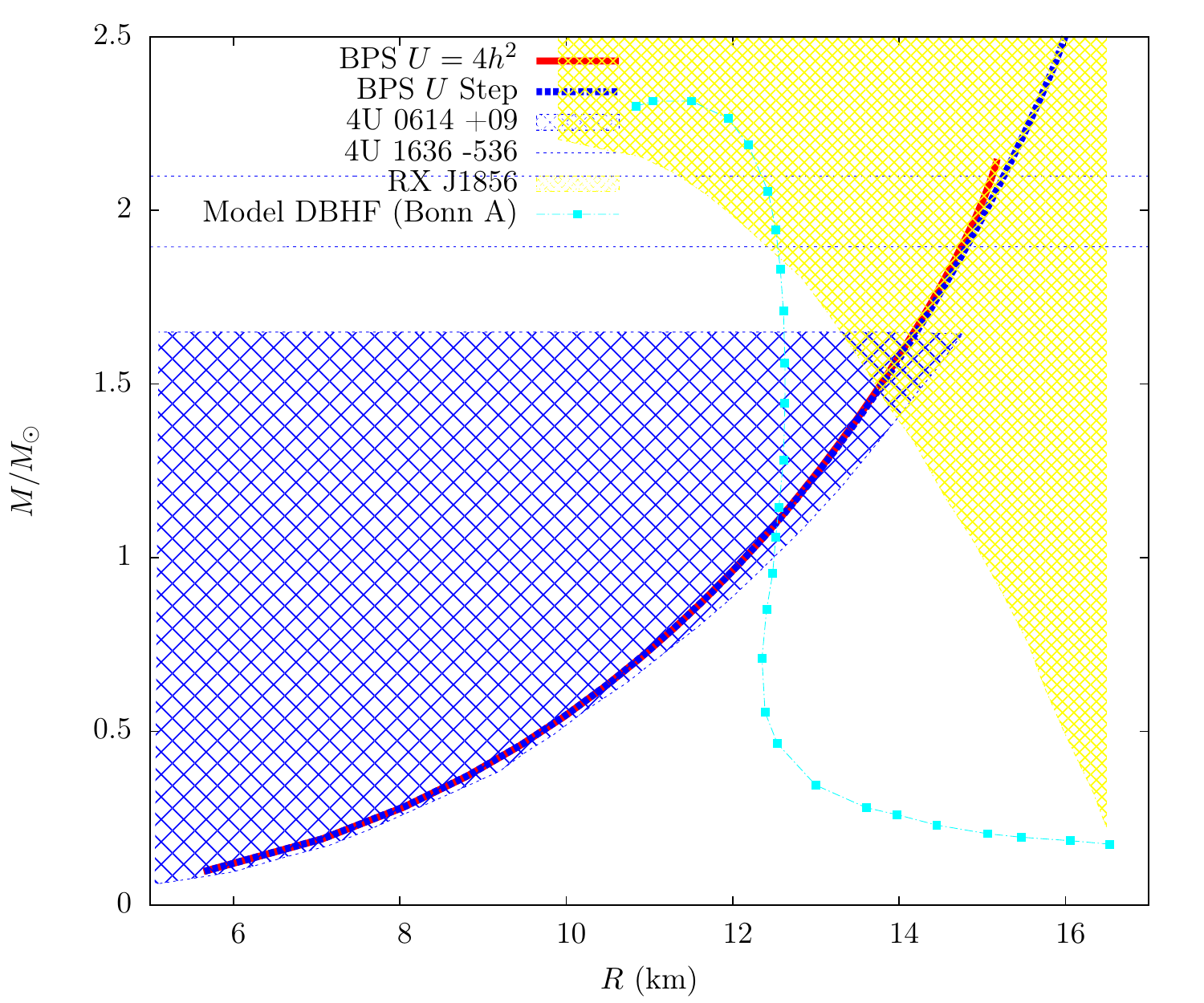}
\caption{Constraints in the $M$-$R$ plane from Ref. \refcite{blaschke}, together with our full field theory curves, both for the partially flat potential ("Step") and the potential $\mathcal{U}_\pi^2$, and the curve for the EoS of the DBHF model. }
\label{constraints}
\end{center}
\end{figure}

\end{itemize}

\section{Summary}\label{sum}
It was the main purpose of this contribution to review the increasing evidence for the relevance of the near-BPS Skyrme model (\ref{full}) as a realistic EFT for nuclei, nuclear matter and some further aspects of low-energy strong interaction physics. This possible relevance is based on several unique properties of the BPS submodel (\ref{BPSmodel}), especially its BPS property, its infinitely many symmetries, and the fact that it is a perfect fluid. The BPS property has the potential to solve the problem of the too high binding energies of the standard Skyrme model, as is already indicated by the results of our (mainly analytical) calculations. The plasticity of BPS skyrmions related to the infinitely many symmetries probably allows to maintain the successes of the standard Skyrme model in the calculations of spin/isospin excitational spectra. A final detailed and quantitative confirmation of both of these assumptions, however, will require a further development of either numerical or perturbative methods in the treatment of the full near-BPS model. 

The perfect fluid property of the BPS submodel not only reflects qualitative properties of nuclei and nuclear matter but, in addition, allows to derive thermodynamical and fluid mechanical properties of skyrmionic matter directly on a microscopical basis, without the necessity to employ a thermodynamical or MF limit. Still, the model permits a simple MF limit and, consequently, both a microscopic (full field theoretic) and a macroscopic (MF) description. This fact, in particular, facilitates the direct comparison with other models of nuclear matter, from which the inevitability of the sextic term $\mathcal{L}_6$ in (\ref{BPSmodel}) immediately follows. Indeed, the effect of this term on nuclear matter is equivalent to the repulsive force of the $\omega$ meson or of the (quark-) current-current interaction, which dominates the behaviour (e.g., the EoS) of nuclear matter at high densities.    

The BPS submodel also allows for a simple description of neutron stars. We only used the classical soliton solutions for this purpose, corresponding to infinite nuclear matter. The resulting fully self-gravitating field configurations result in a rather stiff skyrmionic matter above nuclear saturation density and, therefore, in a gravitational compression which is less pronounced than in most other neutron star models. This leads to mass-radius curves where mostly $dM/dR >0$ (except probably very close to the maximum mass). The $M(R)$ curves are, in fact, qualitatively similar to the $M(R)$ curves of quark stars, allowing, however, for significantly higher maximum NS masses compatible with all existing observational data. 

In the case of neutron stars, the existence of a MF limit allows to use the resulting MF EoS for a TOV type calculation of neutron stars in the BPS Skyrme model, and for a comparison of the full field theory and MF TOV calculations. It turns out that the differences between the two can be considerable, which has important repercussions for the physics of neutron stars, even beyond Skyrme models. 
The inhomogeneities in the BPS Skyrme model and for the used ansatz are, in fact, quite mild and only visible at large distances. In models of nuclear matter with short-distance inhomogeneities (like, e.g., the Skyrme crystal, or models with inhomogeneous condensates\cite{bub}) these problems most likely become more acute.
In particular, this casts some doubts on the reliability of the so-called TOV inversion (the reconstruction of the nuclear matter EoS from the $M(R)$ curve of neutron stars). 

Further improvements in the description of neutron stars within the BPS model include the introduction of a difference between neutrons and protons either in-medium (via the introduction of an isospin chemical potential\cite{iso-chem}) or in-vacuum (via the collective coordinate quantization), or the combination of the BPS model at high densities with established phenomenological EoS of nuclear matter at lower densities. Also the generalization to the full near-BPS model should lead to an  improved description, especially for the NS crust, where, however, either a macroscopic treatment of the resulting skyrmionic matter (e.g., a Skyrme crystal, or a kind of MF approximation) or a full numerical calculation is required, because a simple ansatz leading to spherically symmetric densities in the full field theory is not available in this model. 

In any case, it is our hope that the results and ideas presented here, together with some recent important progress in other variants of the Skyrme model (see, e.g., Refs. \refcite{lau,SutBPS,BPSM,nitta},\refcite{krusch,scattering, hab-lau,HLS,he1,he2,hal}) may finally pave the way for a renewed effort in the further evolution of the old ideas of Skyrme, where the well-known qualitative successes of the Skyrme model are complemented by precise quantitative calculations of properties of nuclei and nuclear matter. In the case of the near-BPS Skyrme model, this requires a detailed program of numerical investigation as well as the development of the necessary numerical tools, as already pointed out in several occasions in the present contribution. 

\section*{Acknowledgements}
The authors acknowledge financial support from the Ministry of Education, Culture, and Sports, Spain (Grant No. FPA 2014-58-293-C2-1-P), the Xunta de Galicia (Grant No. INCITE09.296.035PR and Conselleria de Educacion), the Spanish Consolider-Ingenio 2010 Programme CPAN (CSD2007-00042), and FEDER. 
CN thanks the Spanish Ministery of
Education, Culture and Sports for financial support (grant FPU AP2010-5772). Further, the authors 
thank T. Kl\"ahn for helpful comments.

\bibliographystyle{ws-rv-van}
\bibliography{ws-rv-sample}

\end{document}